# Lattice dynamics of *Pnma* Sn(S$_{1-x}$Se$_x$) solid solutions: energetics, phonon spectra and thermal transport

Jonathan M. Skelton[1]*

[1] *Department of Chemistry, University of Manchester, Oxford Road, Manchester M13 9PL, UK*

* Corresponding author - e-mail jonathan.skelton@manchester.ac.uk

**Abstract**

Alloying is widely used as a means to fine-tune the properties of thermoelectric materials by reducing the lattice thermal conductivity. However, the effects of compositional variation on the lattice dynamics of alloy systems are not well understood, due in part to the difficulty of building realistic first-principles models of structurally-complex solid solutions. This work builds on our previous study of Sn$_n$(S$_{1-x}$Se$_x$)$_m$ solid solutions [Gunn *et al.*, *Chem. Mater.* **31**, *10*, 3672, **2019**] to explore the lattice dynamics of the *Pnma* Sn(S$_{1-x}$Se$_x$) system, which has been widely studied for potential thermoelectric applications. We find that the vibrational internal energy and entropy have a large quantitative impact on the mixing free energy and are likely to be particularly important in alloy systems with competing phases. The thermodynamically-averaged phonon dispersions and density of states curves show that alloying preserves the structure of the low-frequency bands of modes associated with the Sn sublattice but broadens the high-frequency chalcogen bands into a near-continuous spectrum at the 50/50 mixed composition. This results in a general reduction in the phonon mode group velocities and an increase in the number of energy-conserving scattering channels for heat-carrying low-frequency modes, which is consistent with the decrease in thermal conductivity observed in experimental measurements. Finally, we discuss some of the limitations of our first-principles modelling approach and propose methods to address these in future studies.

**Introduction**

Around two thirds of the energy used globally is wasted as low-grade heat from sources including industry, refrigeration and transportation[1]. Thermoelectric (TE) power provides a means to recover electrical energy from temperature gradients and thereby enable significant efficiency gains in these energy-intensive processes, making it an important complement to primary renewable energy sources such as photovoltaics (PV). In particular, in the drive to meet ambitious climate change targets, emissions from internal combustion engines are becoming a huge area of concern, and employing thermoelectric generation as part of a mild hybrid powertrain could provide an interim solution while electric vehicle technology and infrastructure matures to allow for more widespread adoption.

The efficiency of a thermoelectric material is commonly expressed by the dimensionless figure of merit $ZT$[2]:

$$ZT = \frac{S^2 \sigma T}{\kappa_{\text{latt}} + \kappa_{\text{el}}} \quad (1)$$

The Seebeck coefficient $S$, electrical conductivity $\sigma$ and electronic thermal conductivity $\kappa_{\text{el}}$ are interdependent such that the balance of a high power factor $S^2\sigma$ and low $\kappa_{\text{el}}$ is generally found in doped semiconductors. At low to moderate temperatures, the lattice thermal conductivity $\kappa_{\text{latt}}$ makes the most significant contribution to the denominator, and thus minimising $\kappa_{\text{latt}}$ has proven to be a facile means of improving TE performance. In some cases the electronic and thermal transport can be almost completely decoupled[1], as in the "phonon glass, electron crystal" concept[3]. All four parameters in Eq. 1 are implicit functions of temperature, requiring that materials are either optimised to produce a high peak $ZT$ at a target operating temperature range or tuned to display a high $ZT$ across a wide range of temperatures.

Current flagship TE materials include PbTe, SnSe and $Bi_2Te_3$, all three of which display favourable electrical properties and intrinsically low thermal conductivity due to a combination of heavy elements and strongly anharmonic lattice dynamics[4–6]. However, PbTe and $Bi_2Te_3$ are not suitable for widespread adoption due to the low abundance of Te, and there are also concerns with SnSe due to the environmental toxicity of Se. There has therefore been significant effort devoted to alternative systems including the more earth-abundant SnS[2] and metal oxides[7–10].

Alloying is commonly used as a means to enhance thermoelectric performance, as a suitable choice of components can maintain or improve a favourable electronic structure while reducing $\kappa_{\text{latt}}$ by introducing variation in atomic masses and chemical bond strength to promote stronger phonon scattering[11]. Pb(S,Se,Te), Sn(S,Se) and $(Bi,Sb)_2(Se,Te)_3$ alloys have all been studied as thermoelectrics and the alloying shown to improve $ZT$[12–17]. Due to the record-breaking high-temperature $ZT$ of SnS, the Te-free Sn(S,Se) alloys are of particular interest, and experimental studies have demonstrated a reduction in the thermal conductivity of mixed compositions[12,14].

Thermoelectricity is unique in that all four of the terms in the figure of merit $ZT$ are amenable to calculation using first-principles electronic-structure methods such as density-functional theory[18]. There have been a number of theoretical studies on existing and potential new single-component bulk thermoelectrics[6,19–23], and there have recently been efforts to use high-throughput modelling to screen large numbers of compounds in a bid to identify new candidate thermoelectrics[24,25].

Studying multicomponent alloy systems with first-principles methods is challenging, however, as building a realistic model requires calculations on a large number of configurations that for complex systems with low-symmetry unit cells can easily approach the cost of a screening study[26]. Ektarawong and Alling used a first-principles cluster expansion method to study the *Pnma* Sn(S$_{1-x}$Se$_x$) system and confirmed the stability of the mixed phases at finite temperature[27]. We recently carried out a study of four Sn$_n$(S$_{1-x}$Se$_x$)$_m$ alloys[26], including the *Pnma* and rocksalt monosulphides, by performing first-principles calculations on the full sets of symmetry-inequivalent structures formed by substituting chalcogen atoms in small supercells of the parent structures[28]. In this study we also confirmed the stability of the mixed *Pnma* Sn(S$_{1-x}$Se$_x$) alloys and further predicted the effect of alloying on the structural, electrical and optical properties[26].

While many large-scale modelling studies make use of (semi-)local DFT methods to evaluate and compare energetics and electrical properties[29–32], modelling lattice dynamics and in particular thermal transport can be much more resource intensive, and there are thus comparatively fewer examples of screening studies focussing on these properties[33]. As a result, despite the widespread use of alloying to tune the performance of thermoelectric materials, there has been little theoretical investigation into how composition affects the dynamics and related properties. In this work, the model developed in Ref. [26] is extended to include explicit evaluation of the lattice dynamics of the *Pnma* Sn(S$_{1-x}$Se$_x$) solid solution. We quantify the effect of the phonon free energy on the mixing energetics, establish the effect of alloying on the phonon spectrum and frequency dispersion, and explore some of the implications for the thermal transport. Finally, we discuss the broader applicability of this modelling technique and possible future enhancements.

**Computational modelling**

The starting point for our calculations was the set of optimised structures in the *Pnma* Sn(S$_{1-x}$Se$_x$) solid solution model from our previous study[26]. In our previous work, the Transformer code[34] was used to enumerate the symmetry-independent structures formed by successively substituting the chalcogen atoms in a 32-atom 2×1×2 expansion of the *Pnma* SnS/SnSe structure, yielding a total of 2,446 unique structures across seventeen compositions from $x_{Se}$ = 0 (SnS) to $x_{Se}$ = 1 (SnSe). In the present study, lattice dynamics calculations were performed on a subset of 1,294 structures in nine compositions, *viz.* $x_{Se}$ = 0, 0.125, 0.25, 0.375, 0.5, 0.625, 0.75, 0.875 and 1, using the finite-displacement method implemented in the Phonopy package.[35]

The accuracy of the phonon frequencies obtained using the finite-displacement method depends on the range of the real-space interatomic force constants (IFCs) evaluated with the chosen supercell expansion. Taking into account the symmetry of individual structures in the alloy model, where present, between 8 and 192 accurate calculations were required to evaluate the IFCs for each structure, generally representing a considerably larger computational workload than the initial geometry optimisations. For this reason, the force constants were calculated using the 32-atom alloy supercell structures, i.e. without further expansion.

Phonon density of states (DoS) and atom-projected DoS (PDoS) curves were evaluated using Fourier interpolation to obtain frequencies on uniform Γ-centered grids of phonon wavevectors with 24×16×24 subdivisions, with a Gaussian smearing of width $\sigma$ = 0.032 THz (FWHM ~2.5 cm$^{-1}$). The vibrational contributions to the thermodynamic free energy were evaluated using denser sampling meshes with 36×24×36 subdivisions. Phonon band dispersions were evaluated using the band-unfolding scheme implemented in Phonopy[36] to project the vibrational modes calculated for the alloy supercells back onto the parent *Pnma* structure. The dispersion curves were evaluated across a series of segments passing through all the high-symmetry wavevectors in the *Pnma* Brillouin zone. Calculations of the lattice thermal conductivity of the SnS and SnSe endpoints, and evaluation of the two-phonon joint density of states of the alloy models, were performed using the Phono3py package.[37]

All calculations were performed using pseudopotential plane-wave density-functional theory (DFT) as implemented in the Vienna *Ab initio* Simulation Package (VASP) code[38]. As in our previous study[26], a plane-wave cutoff of 600 eV was employed alongside Monhkhorst-Pack *k*-point sampling grids[39] with 4×4×4 subdivisions. The PBEsol exchange-correlation functional with the DFT-D3 dispersion correction was used with projector augmented-wave (PAW) pseudopotentials[40,41] including the Sn 5s, 5p and 4d and the outermost 3d/3p and 4s/4p electrons on S and Se in the valence shell. The precision of the charge-density grids was set automatically to avoid aliasing errors, and a support grid with 8× as many points was used to evaluate the forces. The PAW projection was performed in reciprocal space and nonspherical contributions to the gradient corrections inside the PAW spheres were accounted for.

**Results and Discussion**

*a. Mixing thermodynamics*

The solid solution model for a given composition $x_{\text{Se}}$ comprises a set of $N$ optimised structures with total energy $E_n$ and degeneracy $g_n$, which are used to form the thermodynamic partition function at a temperature $T$, $Z(x_{\text{Se}}, T)$, according to[28]:

$$Z(x_{\text{Se}}, T) = \sum_{n=1}^{N} g_n \exp(-E_n/k_\text{B} T) \quad (2)$$

where $k_\text{B}$ is the Boltzmann constant. $Z(x_{\text{Se}}, T)$ is then used to obtain the constant-volume (Helmholtz) free energy $A = U - TS$ from the bridge relation:

$$A(x_{\text{Se}}, T) = -k_\text{B} T \ln Z(x_{\text{Se}}, T) \quad (3)$$

From $A(x_{\text{Se}}, T)$, the free energy of mixing $A_{\text{mix}}(x_{\text{Se}}, T)$ can be obtained as:

$$A_{\text{mix}}(x_{\text{Se}}, T) = A(x_{\text{Se}}, T) - [(1 - x_{\text{Se}}) A(x_{\text{Se}} = 0, T) + x_{\text{Se}} A(x_{\text{Se}} = 1, T)] \quad (4)$$

where $A(x_{\text{Se}} = 0, T)$ and $A(x_{\text{Se}} = 1, T)$ are the free energies of the SnS and SnSe endmembers respectively, which, since $g_n = 1$, are simply equal to the corresponding $E_n$.

For a given composition and temperature the occurrence probability of individual structures $P_n$ in the model are calculated from the partition function as:

$$P_n(T) = \frac{1}{Z(T)} g_n \exp(-E_n/k_\text{B} T) \quad (5)$$

The $P_n$ can then be used to calculate the thermodynamic average $\bar{X}$ of a general property $X$ using:

$$\bar{X}(T) = \sum_{n=1}^{N} P_n(T) X_n \tag{6}$$

where $X_n$ are the properties of the individual structures in the model.

This model can be improved upon by using the phonon frequencies obtained from the lattice-dynamics calculations to calculate for each structure the temperature-dependent Helmholtz energy $A_n$ including contributions from vibrations:

$$A_n(T) = E_n + A_n^{\text{vib}} = E_n + U_n^{\text{vib}}(T) - T S_n^{\text{vib}}(T) \tag{7}$$

where $A_n^{\text{vib}}$ is the vibrational Helmholtz energy and $U_n^{\text{vib}}$ and $S_n^{\text{vib}}$ are the vibrational internal energy and entropy. The superscript vib is used for clarity to explicitly identify quantities calculated based on the lattice dynamics. As for the free energy of the solid solution, $A_n^{\text{vib}}(T)$ is calculable from the vibrational partition function $Z_n^{\text{vib}}(T)$ using the bridge relation:

$$Z_n^{\text{vib}}(T) = \prod_{\mathbf{q}\nu} \frac{\exp(-\hbar\omega_{\mathbf{q}\nu}/2k_\text{B}T)}{1 - \exp(-\hbar\omega_{\mathbf{q}\nu}/k_\text{B}T)} \tag{8}$$

$$A_n^{\text{vib}}(T) = -\frac{1}{N} k_\text{B} T \ln Z_n^{\text{vib}}(T)$$
$$= \frac{1}{N}\left\{ \frac{1}{2}\sum_{\mathbf{q}\nu} \hbar\omega_{\mathbf{q}\nu} + k_\text{B} T \sum_{\mathbf{q}\nu} \ln[1 - \exp(-\hbar\omega_{\mathbf{q}\nu}/k_\text{B}T)] \right\} \tag{9}$$

Here the phonon frequencies $\omega$ are indexed by a phonon wavevector $\mathbf{q}$ and band index $\nu$ and $\hbar$ is the reduced Planck constant. The product in Eq. 8 runs over all bands on a grid of $N$ wavevectors sampling the phonon Brillouin zone, which appears as a normalisation constant in Eq. 9. The expansion in Eq. 9 shows that $A_n^{\text{vib}}$ is a sum of the harmonic zero-point energies $\frac{1}{2}\hbar\omega_{\mathbf{q}\nu}$ and a temperature-dependent term from the population of higher mode energy levels at finite temperature.

We note that this model assumes the distribution of the different supercell configurations is in thermodynamic equilibrium at the formation temperature $T_\text{F}$[28], and that the chosen alloy supercell is sufficient to sample the range of local configurations that may be present. In real alloy systems, there may be medium- or long-range deviations in local structure and/or composition that are not probed by this method, for example due to the flexible oxidation state and lone-pair activity of Sn[42].

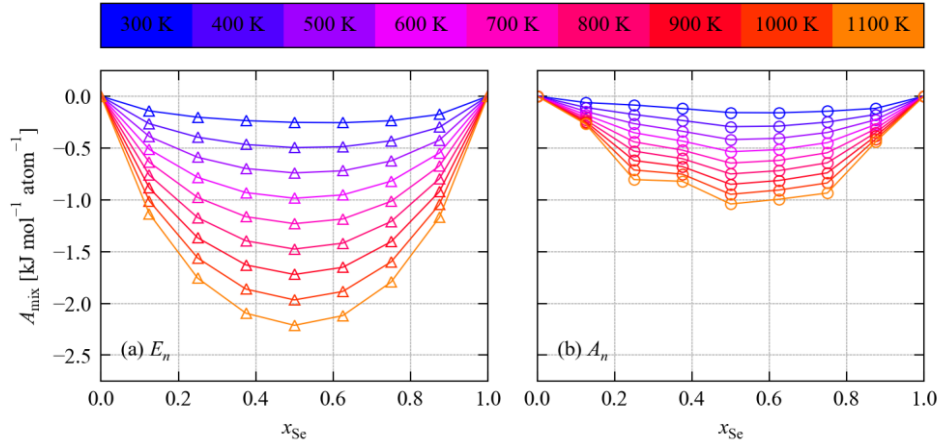

**Figure 1** Calculated mixing free energies $A_{\mathrm{mix}}$ of *Pnma* Sn(S$_{1-x}$Se$_x$) solid solutions as a function of Se fraction $x_{\mathrm{Se}}$, obtained using the lattice energies $E_n$ (a) and the Helmholtz free energies $A_n$ (b) including contributions from the lattice dynamics. Each plot shows $A_{\mathrm{mix}}(x_{\mathrm{Se}})$ for a range of alloy formation temperatures $T_F$ from 300 - 1100 K, which are denoted by line colours from blue (low $T_F$) to orange (high $T_F$).

By substituting the $E_n$ in Eq. 2 with $A_n$ (Eq. 7), we can build the partition function for the solid solution and evaluate the mixing free energies including the effect of the lattice dynamics (Fig. 1). Interestingly, we find that including the vibrational free energy terms destabilises the solid solution, decreasing $A_{\mathrm{mix}}$ for the 50/50 composition ($x_{\mathrm{Se}}$ = 0.5) by a factor of two from -1.72 to -0.85 kJ mol$^{-1}$ per atom at 900 K compared to using the lattice energies. This difference corresponds to $\sim 0.1 \times k_B T$, which is in the typical range of vibrational entropy differences observed by van de Walle and Ceder[43].

For an ideal solid solution, the mixing is purely entropic and the $A_{\mathrm{mix}}$ profile would be symmetric with a minimum at $x_{\mathrm{Se}}$ = 0.5. The $A_{\mathrm{mix}}$ profiles obtained using the $E_n$ show near-ideal behaviour, with the mixing energies of the $x_{\mathrm{Se}}$ = 0.25 and $x_{\mathrm{Se}}$ = 0.75 compositions differing by 0.04 kJ mol$^{-1}$ atom$^{-1}$ at 900 K. However, a much larger difference of 0.12 kJ mol$^{-1}$ atom$^{-1}$ is obtained when including the free energy, indicating a larger deviation from ideality, which is clearly evident when comparing the two sets of mixing profiles in Fig. 1.

Of particular note are the large deviations from the general trend in Fig. 1a at $x_{\mathrm{Se}}$ = 0.125 and 0.375, the reasons for which are not clear. One possible explanation is that the 32-atom supercell does not adequately capture the full range of local structures present in the bulk alloy, although in our previous study we found that this was at least sufficient to converge the mixing energies calculated from the $E_n$[26]. Another possibility is that the deviation is an artefact of the limited real-space range of the calculated force constants and its impact on the interpolation of the phonon frequencies to a dense grid of wavevectors to evaluate $A_n^{\mathrm{vib}}$. Although we would naively expect a similar level of accuracy across the nine compositions, we found that the proportion of imaginary frequencies in the wavevector sampling meshes - which are excluded from the partition function in Eq. 8 - ranged from 2.06 $\pm$ 0.91 % for the $x_{\mathrm{Se}}$ = 0.875 composition to 2.81 $\pm$ 0.6 % for $x_{\mathrm{Se}}$ = 0.375, so this is a possibility. As an additional test we also recalculated the mixing energy based on the $A_n^{\mathrm{vib}}$ obtained with a denser sampling mesh with 54×36×54 subdivisions and obtained quantitatively similar results (Fig. S1), with a maximum difference of 14 $\mu$eV in the absolute free energy per atom across the nine compositions.

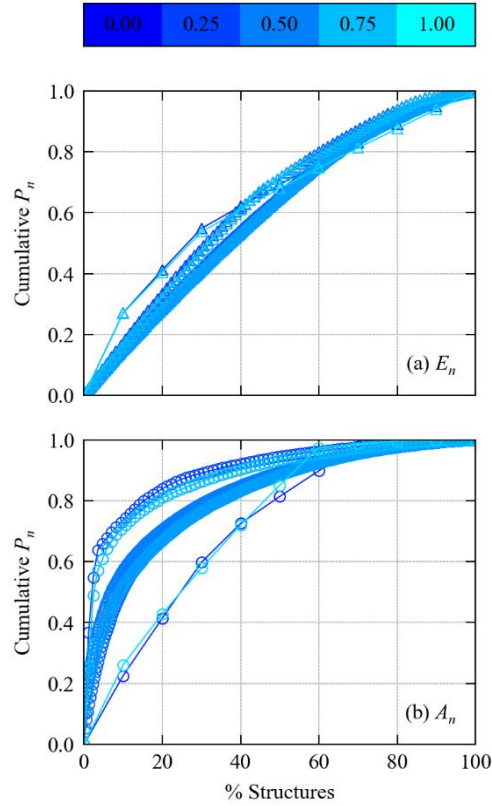

**Figure 2** Cumulative occurrence probability $P_n$ as a function of the percentage of structures in the solid solution models with compositions ranging from $x_{\text{Se}}$ = 0.125 to 0.875, calculated using the lattice energies $E_n$ (a) and the free energies $A_n$ (b).

In both cases, the mixing is still energetically favoured at all compositions and the qualitative stability of the alloy is thus unaffected. However, the substantial difference in the mixing energy may be important in systems with competing phases. In our previous work[26,42], we predicted based on lattice energies that the energy difference between competing *Pnma* and rocksalt $Sn(S_{1-x}Se_x)$ monochalcogenide solid solutions varies from 10.5 and 2.6 kJ mol$^{-1}$ per F.U. between the SnS and SnSe endpoints[26]. We also found that the vibrational free energy stabilises the rocksalt phase of SnS relative to the *Pnma* phase at higher temperatures, reducing the energy difference to 4.5 kJ mol$^{-1}$ per F.U. at 900 K[42]. It is therefore conceivable that differences in the free energy may bring the rocksalt phase closer to the convex hull at Se-rich compositions. To test this would require an additional set of lattice dynamics calculations on the rocksalt solid solution models, which we have not attempted here.

Including the vibrational free energy in the partition function also leads to a marked change in the distribution of the occurrence probabilities of individual structures. For the 50/50 composition ($x_{\text{Se}}$ = 0.5) and a formation temperature $T_\text{F}$ of 900 K, the $P_n$ calculated using the lattice energies range from 8.8 × 10$^{-5}$ to 4.2 × 10$^{-3}$, whereas including the effect of lattice dynamics produces a significantly wider spread of 2.3 × 10$^{-5}$ - 5.2 × 10$^{-2}$. As shown in Fig. 2, the discrepancy is such that 20 % of structures with the highest $P_n$ account for > 70 % of the partition function formed using the free energies, compared to just under 30 % with the lattice energies.

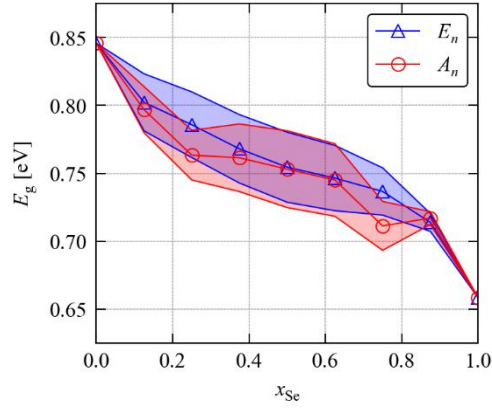

**Figure 3** Composition dependence of the electronic bandgap $E_g$ calculated using occurrence probabilities $P_n$ based on the lattice energies $E_n$ (blue triangles) and the Helmholtz energies $A_n$ (red circles). The markers show the calculated averages and the shaded regions show ± one weighted standard deviation.

The vibrational free energy is derived from the phonon frequencies, which in this system are much more sensitive to the structure than the lattice energy. Again taking the $x_{Se}$ = 0.5 model as an example, the largest difference in $E_n$ between two structures is 3.8 meV atom$^{-1}$ (0.36 kJ mol$^{-1}$ atom$^{-1}$), whereas the difference in the zero point energies is 0.06 kJ mol$^{-1}$ atom$^{-1}$ and the difference in $A_n^{\mathrm{vib}}$ at 900 K is 2.06 kJ mol$^{-1}$ atom$^{-1}$. Given the number of examples in the literature showing that differences in vibrational free energy - in particular the vibrational entropy - are a key driver of temperature-induced phase transitions[42,44–46] and are important in the determining the stability of alloy systems[43,47,48] - this finding is not surprising, but is nonetheless noteworthy.

In principle, the large change in $P_n$ may lead to a significant change in the averaged properties, although this is likely to be mitigated if the variation in properties among the structures is small. To test this, we took the electronic bandgaps of the *Pnma* structures obtained in our previous study and calculated the averaged bandgap $\bar{E}_g$ as a function of composition using the 900 K occurrence probabilities calculated using the $E_n$ and $A_n$ (Eq. 6, Fig. 3). The two sets of calculated bandgaps for the 50/50 composition differ by only ~1.5 meV, and including the vibrational free energy reduces the averaged bandgaps of the $x_{Se}$ = 0.25 and $x_{Se}$ = 0.75 compositions by 2.8 and 3.4 %, respectively, from 0.79 and 0.74 eV to 0.76 and 0.71 eV. These reductions may be compared to the range of $E_g$ of the individual structures in the three compositions, which are 111, 212 and 80 meV. The weighted standard deviations are also largely unchanged, with the largest difference being a decrease of 5.9 meV for the $x_{Se}$ = 0.25 composition. Our calculations thus suggest that the differences in the $P_n$ do not have a significant impact on the predicted electrical properties of the *Pnma* Sn(S$_{1-x}$Se$_x$) alloy.

Finally, we note that in this analysis we have used the Helmholtz rather than the Gibbs free energy $G(x_{Se}, T)$ as in our previous study. In the absence of contributions to the free energy from the lattice dynamics, the effect of finite pressure can be modelled by replacing the internal energies $E_n$ with the enthalpies $H_n = E_n + pV_n$[49]. This may be done approximately by adding a correction $pV_n$ to each $E_n$ or by relaxing the structures under applied pressure to evaluate the $H_n$ explicitly. However, thermal expansion at finite temperature has a considerable effect on the phonon spectra and the vibrational contributions to the free energy, which should be accounted for by calculating $G_n$ for each structure within the quasi-harmonic approximation (QHA)[35]. However, to do so for all of the structures in the solid solution model would require an order of magnitude more calculations and is therefore not feasible using first-principles techniques.

*b. Phonon spectra*

For a given composition and formation temperature $T_F$, a thermodynamically averaged phonon density of states $g(v)$ (DoS) can be obtained by summing the DoS curves of individual structures weighted by the corresponding occurrence probabilities $P_n(T_F)$ (Eq. 5/Eq. 6). Obtaining an averaged phonon dispersion is more involved, as the vibrational modes calculated in the alloy supercells must first be "unfolded" onto the parent structure. In the present study we employed the method proposed by Allen *et al.*[36], which is implemented in recent versions of the Phonopy lattice-dynamics code[35] and accessible through the Python API. Reference structures for the unfolding were generated for each composition by linear interpolation of the lattice vectors and atomic positions between the SnS and SnSe structures.

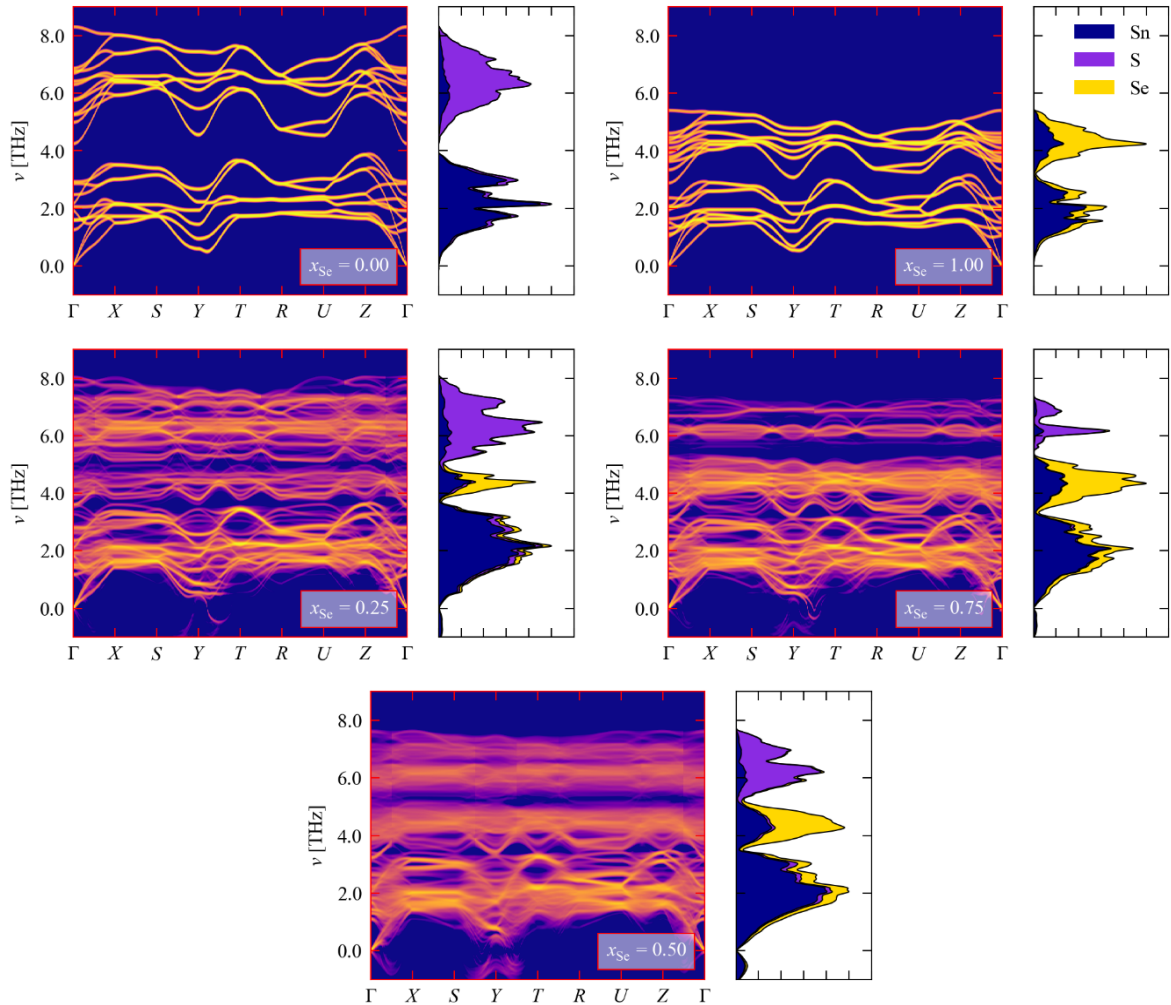

**Figure 4** Phonon spectra of *Pnma* Sn(S$_{1-x}$Se$_x$) solid solutions with composition $x_{Se}$ = 0, 0.25, 0.5, 0.75 and 1. Each subplot shows the dispersion of the alloy supercells unfolded back to a *Pnma* parent structure and the density of states $g(v)$ (DoS) as a stacked area plot showing the projection onto Sn (blue), S (purple) and Se atoms (yellow). For the three intermediate compositions, the averaging over structures in the solid-solution model was performed based on a formation temperature $T_F$ of 900 K.

Fig. 4 compares the dispersions and atom-projected DoS curves of the SnS and SnSe endpoints to those of three intermediate compositions with $x_{Se}$ = 0.25, 0.5 and 0.75. The phonon spectra for the intermediate compositions in Figs. 4c-4e were averaged assuming a 900 K formation temperature (there is only one unique structure for each endpoint, so no averaging is required to generate the spectra in Figs. 4a/4b). A full set of phonon spectra for the nine compositions can be found in Figs. S2-S10.

The spectra of the endpoints each comprise two bands of modes primarily associated with motion of the Sn and chalcogen atoms. At intermediate compositions, the structure of the low-frequency Sn peak is largely unchanged, while the higher-frequency chalcogen feature splits into a mid-frequency Se and high-frequency S component. To a large extent, the Sn sublattice in the mixed phases retains the parent *Pnma* structure, and this is reflected in the low-frequency part of the unfolded dispersions retaining significant detail. On the other hand, substituting the chalcogen atoms spreads out the mid- and high-frequency parts of the dispersion, leading to almost continuous bands in the 50/50 composition with $x_{Se}$ = 0.5.

The broadening of the frequency spectrum at intermediate compositions may have significant implications for the thermal transport, as it could both decrease the mode group velocities and increase the density of energy-conserving phonon scattering events and thereby suppress the mode lifetimes. These points will be returned to in the following subsection.

It can also be seen from the unfolded dispersions and DoS curves that some of the structures in the mixed compositions display imaginary modes. As noted in the previous section, the force constants were calculated to a limited real-space range, and the unfolded dispersions and DoS curves are derived by interpolating the zone-centre (Γ-point) modes of the alloy supercells. None of the structures in the solid-solution model show imaginary modes at the zone centre, which implies that these imaginary modes may indeed be interpolation artefacts. However, evaluating the IFCs in an expanded cell would not be practical given the number of structures in the solid-solution model.

Finally, we also examined the spread in the averaged phonon DoS curves by overlaying the weighted standard deviations (Figs. S11-S19). As suggested by the dispersions of the intermediate compositions in Fig. 4, the calculations predict the spread in the DoS, in particular the chalcogen bands, to increase substantially towards the 50/50 composition. We also generated a series of DoS meshes using the larger 36×24×36 sampling mesh used to evaluate the vibrational free energies and the tetrahedron method for Brillouin zone integration, which yielded similar overall shapes and predicted spreads (Figs. S20-S28).

*c. Implications for thermal transport*

Within the single-mode relaxation time approximation (RTA) solution to the Boltzmann transport equation, the macroscopic lattice thermal conductivity tensor $\boldsymbol{\kappa}_{\text{latt}}$ can be calculated as a sum of macroscopic contributions $\boldsymbol{\kappa}_\lambda$ from individual phonon modes $\lambda$ according to[37]:

$$\boldsymbol{\kappa}_{\text{latt}} = \frac{1}{NV} \sum_\lambda \boldsymbol{\kappa}_\lambda = \frac{1}{NV} \sum_\lambda C_\lambda \boldsymbol{v}_\lambda \otimes \boldsymbol{v}_\lambda \tau_\lambda \tag{10}$$

where $C_\lambda$ are the modal heat capacities, $\boldsymbol{v}_\lambda$ are the group velocities, $\tau_\lambda$ are the mode lifetimes, and the macroscopic $\boldsymbol{\kappa}_{\text{latt}}$ is normalised for the unit cell volume $V$ and the number of wavevectors $N$ in the mesh used to integrate over the Brillouin zone. The product $v_\lambda \tau_\lambda$ is the phonon mean-free path $\Lambda_\lambda$, which appears in a frequently-used variant of Eq. 10. $C_\lambda$ and $v_\lambda$ are calculated within the harmonic approximation as:

$$C_\lambda = k_\text{B}\left(\frac{\hbar\omega_\lambda}{k_\text{B}T}\right)^2 \frac{\exp[-\hbar\omega_\lambda/k_\text{B}T]}{(\exp[-\hbar\omega_\lambda/k_\text{B}T]-1)^2} \tag{11}$$

$$\boldsymbol{v}_\lambda = \frac{\partial \omega_\lambda}{\partial \mathbf{q}} = \frac{1}{2\omega_\lambda}\langle \boldsymbol{W}_\lambda \left| \frac{\partial D(\mathbf{q})}{\partial \mathbf{q}} \right| \boldsymbol{W}_\lambda \rangle \tag{12}$$

where $\omega_\lambda$ are the mode frequencies, equivalent to the $\omega_{\mathbf{q}\nu}$ that appear in Eqs. 8 and 9, $\boldsymbol{W}_\lambda$ are the corresponding mode eigenvectors, and $D(\mathbf{q})$ is the dynamical matrix for the phonon wavevector $\mathbf{q}$. $\tau_\lambda$ are the phonon lifetimes and are given by the inverse of the phonon linewidths $\Gamma_\lambda$:

$$\tau_\lambda = \frac{1}{2\Gamma_\lambda} \tag{13}$$

$\Gamma_\lambda$ are calculated as the imaginary part of the phonon self-energy, which can be computed perturbatively to third order from:

$$\begin{aligned}\Gamma_\lambda = \frac{18\pi}{\hbar^2} \sum_{\lambda'\lambda''} &|\Phi_{-\lambda\lambda'\lambda''}|^2 \\ &\times \{(n_{\lambda'} + n_{\lambda''} + 1)\delta(\omega_\lambda - \omega_{\lambda'} - \omega_{\lambda''}) \\ &+ (n_{\lambda'} - n_{\lambda''})[\delta(\omega_\lambda + \omega_{\lambda'} - \omega_{\lambda''}) - \delta(\omega_\lambda - \omega_{\lambda'} + \omega_{\lambda''})]\}\end{aligned} \tag{14}$$

$n_\lambda$ are the phonon occupation numbers from the Bose-Einstein distribution:

$$n_\lambda = \frac{1}{\exp[-\hbar\omega_\lambda/k_\text{B}T]-1} \tag{15}$$

$\Phi_{\lambda\lambda'\lambda''}$ are the three-phonon interaction strengths calculated from:

$$\Phi_{\lambda\lambda'\lambda''} = \frac{1}{\sqrt{N}} \frac{1}{3!} \sum_{\kappa\kappa'\kappa''} \sum_{\alpha\beta\gamma} W_\lambda(\kappa,\alpha) W_{\lambda'}(\kappa',\beta) W_{\lambda''}(\kappa'',\gamma)$$
$$\times \sqrt{\frac{\hbar}{2m_\kappa \omega_\lambda}} \sqrt{\frac{\hbar}{2m_{\kappa'} \omega_{\lambda'}}} \sqrt{\frac{\hbar}{2m_{\kappa''} \omega_{\lambda''}}}$$
$$\times \sum_{l'l''} \Phi_{\alpha\beta\gamma}(0\kappa, l'\kappa', l''\kappa'') \times \exp\{i\mathbf{q}' \cdot [\mathbf{r}(l'\kappa') - \mathbf{r}(0\kappa)]\}$$
$$\times \exp\{i\mathbf{q}'' \cdot [\mathbf{r}(l''\kappa'') - \mathbf{r}(0\kappa)]\} \times \exp[i(\mathbf{q} + \mathbf{q}' + \mathbf{q}'') \cdot \mathbf{r}(0\kappa)]$$
$$\times \Delta(\mathbf{q} + \mathbf{q}' + \mathbf{q}'') \quad (16)$$

The indices $\kappa$ label atoms with mass $m_\kappa$ amd $\alpha$, $\beta$ and $\gamma$ label the Cartesian directions. The third-order IFCs $\Phi_{\alpha\beta\gamma}$ are calculated using the finite-displacement method:

$$\Phi_{\alpha\beta\gamma} = \frac{\partial E}{\partial r_\alpha \partial r'_\beta \partial r''_\gamma} = -\frac{\partial F_\alpha}{\partial r'_\beta \partial r''_\gamma} \approx -\frac{\Delta F_\alpha}{\Delta r'_\beta \Delta r''_\gamma} \quad (17)$$

The indices $l$ label crystallographic unit cells, and the functions $\delta$ and $\Delta$ enforce conservation of energy and crystal momentum, respectively.

Due to the larger number of pairwise atomic displacements that need to be considered, calculating the third-order IFCs in Eq. 16 to compute the phonon linewidths is typically 1-2 orders of magnitude more computationally expensive than obtaining the second-order force constants for the harmonic phonon calculation, and represents the bulk of the computational workload when modelling $\kappa_\text{latt}$ within the RTA.

We have previously modelled the thermal conductivity of *Pnma* SnS and SnSe using the RTA[6,50], based on which we obtained room-temperature (300 K) averaged values of 0.74 and 1.28 W m$^{-1}$ K$^{-1}$ respectively. These are in reasonably good agreement with experimental measurements of 1.2 and 0.4-0.7 W m$^{-1}$ K$^{-1}$[2,51]. The higher thermal conductivity of the selenide predicted by these calculations is possibly unexpected given the general decrease in $\kappa_\text{latt}$ with the formula mass[37], but given the difficulty of preparing high-quality bulk single crystals for measurements it is not clear whether this is a real phenomenon or an issue with the calculations[52,53].

Fig. 5 plots for the two structures the averaged modal contributions to the thermal conductivity $\bar{\kappa}_\lambda = \frac{1}{3}\text{Tr}(\boldsymbol{\kappa}_\lambda)$, the modal heat capacities $C_\lambda$, the group velocity norms $|\boldsymbol{v}_\lambda|$ and the phonon lifetimes $\tau_\lambda$. In both systems, the $\bar{\kappa}_\lambda$ span roughly five orders of magnitude. The heat capacities $C_\lambda$ are a shallow function of frequency and vary by ~10$^{-5}$ eV K$^{-1}$ over the 10 THz range of the SnS phonon spectrum. The $\boldsymbol{v}_\lambda$ in both systems vary over two orders of magnitude from approx. 10$^2$ - 10$^4$ ms$^{-1}$ and are generally higher in the sulphide than the selenide, reflecting the stronger chemical bonding. The mode lifetimes vary by around two orders of magnitude from 0.1 - 10 ps in SnS and 0.5 - 50 ps in SnSe. In both compounds, there is a marked difference in the spectrum of $\tau_\lambda$ for the lower- and higher-frequency modes in the phonon spectrum, with the latter being generally shorter lived.

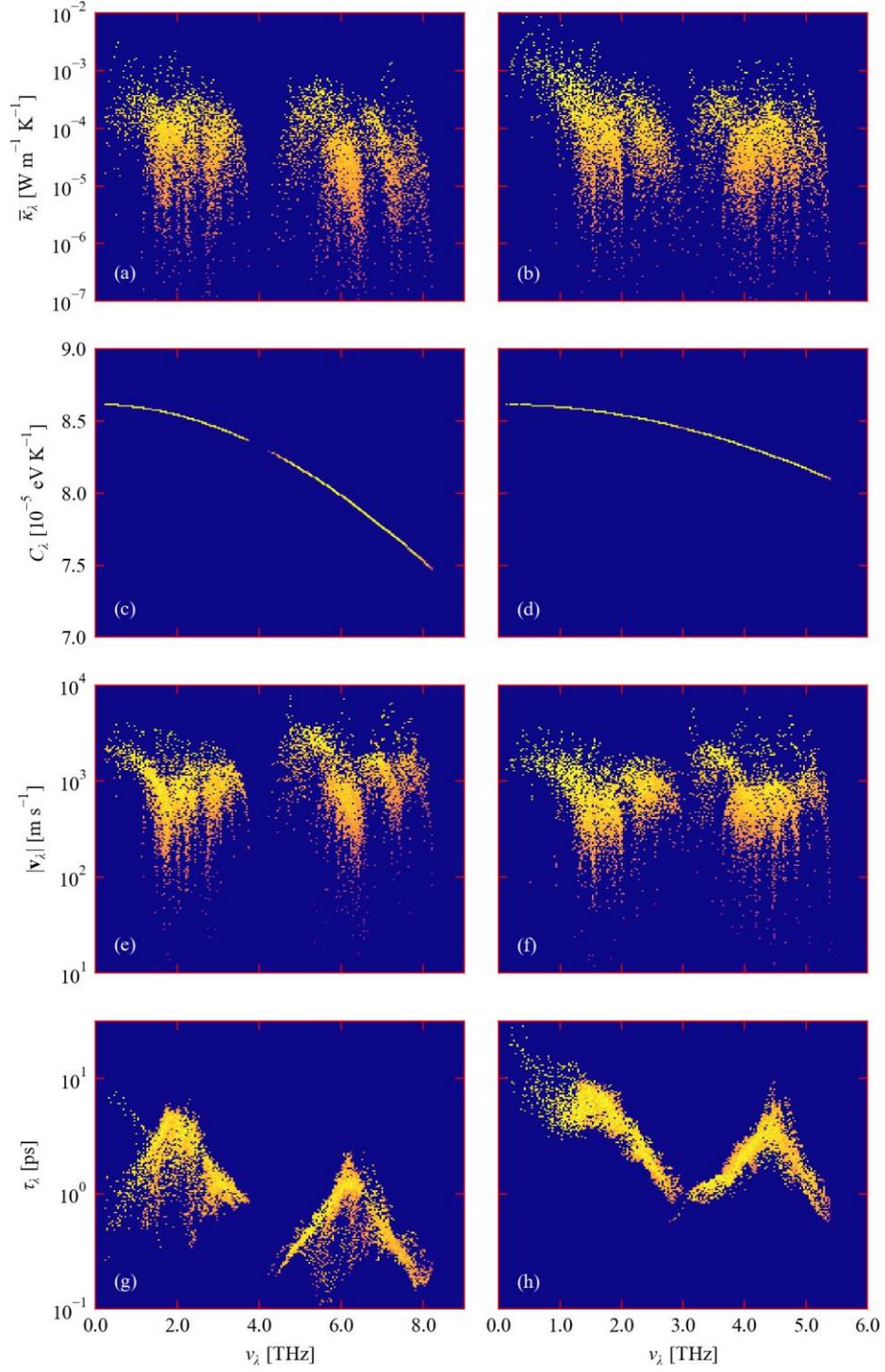

**Figure 5** Frequency dependence of the four modal terms in Eq. 10 calculated for SnS (left column) and SnSe (right column): (a)/(b) - averaged thermal conductivities $\bar{\kappa}_\lambda = \frac{1}{3}\mathrm{Tr}(\boldsymbol{\kappa}_\lambda)$; (c)/(d) - heat capacities $C_\lambda$; (e)/(f) - group velocity norms $|\boldsymbol{v}_\lambda|$; (g)/(h) - lifetimes $\tau_\lambda$. The heat maps are colour coded by $\bar{\kappa}_\lambda$ from red (small $\bar{\kappa}_\lambda$) to yellow (large $\bar{\kappa}_\lambda$). Note that the *y* axes in subplots (c) and (d) are on a linear scale while the other subplots use a logarithmic scale. The data for SnS and SnSe was generated from the calculations in Refs. [50] and [6] respectively.

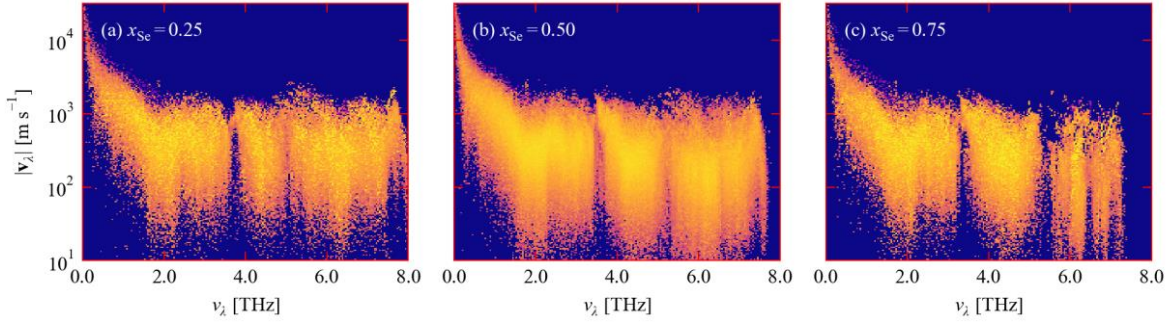

**Figure 6** Frequency spectra of the group velocities in Sn(S$_{1-x}$Se$_x$) solid solutions with composition $x_{Se}$ = 0.25 (a), 0.5 (b) and 0.75 (c). The colour scale indicates the density of modes and runs from red (low density) to yellow (high density). The averaging over structures in the solid-solution model was performed based on a formation temperature $T_F$ of 900 K.

Due to the tensor product $\boldsymbol{v}_\lambda \otimes \boldsymbol{v}_\lambda$ in Eq. 10 the $\bar{\kappa}_\lambda$ are proportional to $v_\lambda^2$ and the spectrum of group velocities largely determines the overall spectrum of $\bar{\kappa}_\lambda$. $\bar{\kappa}_\lambda$ are directly proportional to the to the lifetimes, and the variation in $\tau_\lambda$ with frequency thus imposes additional structure. Comparing SnS and SnSe, it can be seen that the balance of smaller $\boldsymbol{v}_\lambda$ but longer $\tau_\lambda$ in the selenide results in similar predicted overall thermal conductivities. The larger averaged thermal conductivity of the SnSe endpoint seems in particular to arise from a group of low-frequency modes ($v_\lambda$ < 1.5 THz) with group velocities comparable to SnS but with significantly longer lifetimes.

$\boldsymbol{v}_\lambda$ are calculated within the harmonic approximation, making it straightforward to compute an averaged frequency spectrum for three of the mixed compositions in the *Pnma* solid-solution model (Fig. 6). Due to the large number of structures, the $\boldsymbol{v}_\lambda$ of each were generated on a small sampling mesh with 5×3×5 subdivisions. Tests on the SnS and SnSe endpoints using an equivalent 9×3×10 sampling mesh yielded 300 K thermal conductivities to within 5 % of the values obtained with the larger 16×16×16 meshes used in the previous studies[6,50]. Comparing the spectra of the mixed compositions with $x_{Se}$ = 0.25, 0.5 and 0.75 to the pure SnS and SnSe endpoints in Fig. 5 clearly shows that the alloying results in a much broader distribution of $\boldsymbol{v}_\lambda$. The high group velocities of the low-frequency modes are maintained, and the bulk of the modes fall into the same 10$^2$-10$^3$ m s$^{-1}$ range as the endpoints, but there is a general shift to lower $\boldsymbol{v}_\lambda$ among the mid- and high-frequency modes, which becomes more prominent for larger $x_{Se}$. Comparison of the $\boldsymbol{v}_\lambda$ of the full range of compositions (Figs. S29-S37) further highlights the larger spread of group velocities at intermediate compositions, mirroring the broadening of the phonon spectra in Fig. 4.

We note that we are comparing data for the mixed compositions to previous calculations performed without a dispersion correction and using larger supercell expansions to calculate the force constants. In general, the main impact of the exchange-correlation functional on the lattice dynamics is through differences in the predicted equilibrium volume[54]. The optimised volumes of SnS and SnSe obtained in Refs. [6] and [50] are similar to those of the two endpoints in our solid-solution model (46.9/45.8 and 51.8/50.7 Å$^3$ per F.U. respectively), so we do not expect this to be big issue. However, the larger supercell expansion used in the endpoint calculations will in principle make the calculated $\boldsymbol{v}_\lambda$ more accurate by improving the accuracy of the interpolated dispersion and hence the derivatives $\boldsymbol{v}_\lambda = \partial \omega_\lambda / \partial \mathbf{q}$ (c. f. Eq. 12). This may in particular explain why some of the lower-frequency modes in the mixed compositions are predicted to reach higher $|\boldsymbol{v}_\lambda|$ than in either of the endpoints, which is otherwise unexpected.

Due to the significantly higher cost of calculating the third-order IFCs, explicitly computing the lifetimes of even a subset of the solid solution models is impractical. However, it is possible to obtain some qualitative insight into how the alloying may affect the heat transport from the (harmonic) phonon DoS. As noted above, the phonon lifetimes in Eq. 10 are calculated from the phonon linewidths $\Gamma_\lambda$ (Eq. 13), which are in turn computed as a sum of contributions from three-phonon scattering processes with pairs of modes $\lambda'$ and $\lambda''$ (Eq. 14) that depend on the three-phonon interaction strength $|\Phi_{\lambda\lambda'\lambda''}|^2$, the mode occupation numbers and the difference in the phonon frequencies. An approximate linewidth $\tilde{\Gamma}_\lambda$ can thus be written as the product of an averaged phonon interaction strength $P_\lambda$ and a joint density of states $N_2(\mathbf{q}, \omega)$ (JDoS) counting the number of energy-conserving scattering channels as a function of frequency:[37]

$$\tilde{\Gamma}_\lambda = \frac{18\pi}{\hbar^2} P_\lambda N_2(\mathbf{q}_\lambda, \omega_\lambda) \tag{18}$$

The interaction strength $P_\lambda$ is defined as:

$$P_\lambda = \frac{1}{(3n_a)^2} \sum_{\lambda'\lambda''} |\Phi_{\lambda\lambda'\lambda''}|^2 \tag{19}$$

where $n_a$ is the number of atoms in the primitive cell and $3n_a$ is the number of bands at each phonon wavevector. The JDoS $N_2(\mathbf{q}, \omega)$ is a sum of two functions corresponding to collision (Type 1 - two phonons in, one out) and decay processes (Type 2 - one phonon in, two out):

$$N_2(\mathbf{q}_\lambda, \omega_\lambda) = N_2^{(1)}(\mathbf{q}_\lambda, \omega_\lambda) + N_2^{(2)}(\mathbf{q}_\lambda, \omega_\lambda) \tag{20}$$

$N_2^{(1)}(\mathbf{q}_\lambda, \omega_\lambda)$ and $N_2^{(2)}(\mathbf{q}_\lambda, \omega_\lambda)$ are defined as follows:

$$\begin{aligned} N_2^{(1)}(\mathbf{q}_\lambda, \omega_\lambda) = \frac{1}{N} \sum_{\lambda'\lambda''} &\Delta(-\mathbf{q} + \mathbf{q}' + \mathbf{q}'')(n_{\lambda'} - n_{\lambda''}) \\ &\times [\delta(\omega + \omega_{\lambda'} - \omega_{\lambda''}) - \delta(\omega - \omega_{\lambda'} + \omega_{\lambda''})] \end{aligned} \tag{21}$$

$$N_2^{(2)}(\mathbf{q}_\lambda, \omega_\lambda) = \frac{1}{N} \sum_{\lambda'\lambda''} \Delta(-\mathbf{q} + \mathbf{q}' + \mathbf{q}'')(n_{\lambda'} + n_{\lambda''} + 1) \times \delta(\omega - \omega_{\lambda'} - \omega_{\lambda''}) \tag{22}$$

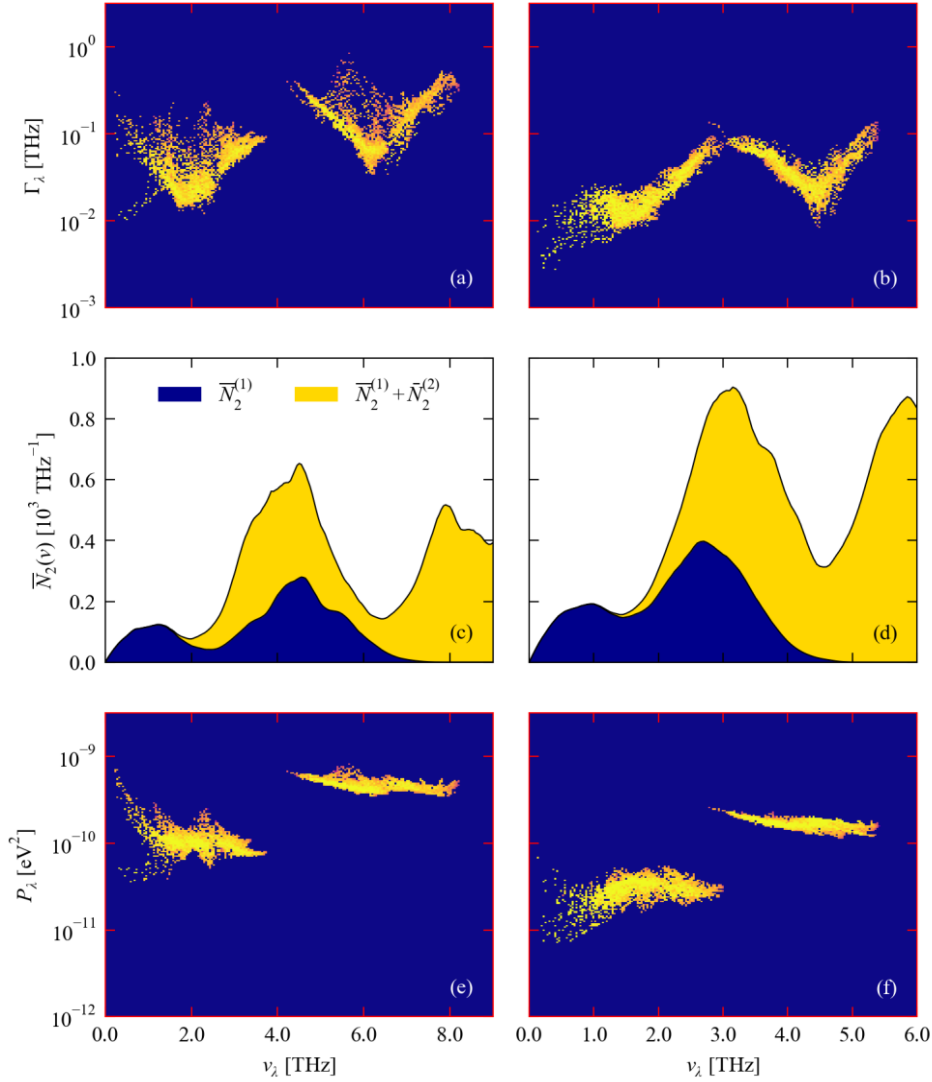

**Figure 7** Analysis of the phonon lifetimes in SnS (left column) and SnSe (right column) using Eqs. 18-23: (a)/(b) - mode linewidths $\Gamma_\lambda$ as a function of frequency; (c)/(d) - two-phonon density of states $\bar{N}_2(\nu)$ shown separately for collision (Type 1 - $\bar{N}_2^{(1)}$) and decay (Type 2 - $\bar{N}_2^{(2)}$) events; (e)/(f) - averaged three-phonon interaction strengths $P_\lambda$. As in Fig. 5 the heat maps are colour coded by $\bar{\kappa}_\lambda$ from red (small $\bar{\kappa}_\lambda$) to yellow (large $\bar{\kappa}_\lambda$). Note that the y-axes in subplots (c) and (d) are on a linear scale while the other subplots use a logarithmic scale. As in Fig. 5 the data for SnS and SnSe was generated from the calculations in Refs. [50] and [6] respectively.

$n_\lambda$ are the phonon occupation numbers at the calculation temperature (Eq. 15) and the functions $\delta$ and $\Delta$ enforce conservation of energy and crystal momentum, respectively (c.f. Eqs. 14/16). Finally, we further define a function $\bar{N}_2(\omega_\lambda)$ and component parts $\bar{N}_2^{(1)}(\omega_\lambda)$ and $\bar{N}_2^{(2)}(\omega_\lambda)$ averaged over wavevectors **q** as follows:

$$\bar{N}_2(\omega_\lambda) = \bar{N}_2^{(1)}(\omega_\lambda) + \bar{N}_2^{(2)}(\omega_\lambda) = \frac{1}{N}\sum_\mathbf{q} N_2^{(1)}(\mathbf{q}_\lambda, \omega_\lambda) + \frac{1}{N}\sum_\mathbf{q} N_2^{(2)}(\mathbf{q}_\lambda, \omega_\lambda) \qquad (23)$$

The JDoS functions are calculable from the harmonic phonon frequencies and, if the $P_\lambda$ can be assumed to be similar in the intermediate phases and endpoints, could provide an indication of the expected composition dependence of the phonon lifetimes. Figure 7 compares the spectrum of phonon linewidths of the SnS and SnSe endpoints against the JDoS functions and the $P_\lambda$. The distribution of $\Gamma_\lambda$ reflects the trends in the lifetimes shown in Fig. 5: in both spectra, the higher-frequency bands of modes show broader linewidths (shorter lifetimes) than the lower-frequency modes, and the sulphide shows broader linewidths than the selenide. Comparison of the linewidth spectra against $\bar{N}_2(\omega_\lambda)$ shows that variation in the density of energy-conserving scattering pathways with frequency is largely responsible for the overall structure, with peaks in $\bar{N}_2(\omega_\lambda)$ overlapping with increases in $\Gamma_\lambda$. Collision (Type 1) and decay (Type 2) events are the dominant scattering processes at low and high frequencies, respectively, while both processes are active at intermediate frequencies. On the other hand, differences in the magnitude of the phonon interaction strengths are clearly responsible for the marked difference in the linewidths of the low- and high-frequency bands of modes, and the lower $P_\lambda$ in the selenide counteract the increase in the JDoS, resulting in narrower linewidths (longer lifetimes) than in the sulphide.

Comparing the $\bar{N}_2(\omega_\lambda)$ of the SnS and SnSe endpoints to averaged JDoS functions calculated for the intermediate compositions (Fig. 8) shows the effect of spreading the phonon DoS over a larger range of frequencies (c.f. Fig. 4). Whereas in the SnS and SnSe endpoints the two JDoS functions $\bar{N}_2^{(1)}$ and $\bar{N}_2^{(2)}$ show distinct peaks due to the well-defined bands of modes in the respective phonon spectra, the JDoS functions of the mixed compositions are much broader, implying stronger scattering across a broad spectrum of mid- and high-frequency modes. However, the calculations predict the JDoS of all three mixed compositions to be on a similar scale to the endpoints, which suggests that the broader range of phonon frequencies in the mixed compositions does not necessarily lead to as large an enhancement in the number of scattering channels as might be expected. This is, however, consistent with the relatively small absolute variation in $\kappa_{\text{latt}}$ observed among different alloy compositions in experiments[12,14].

The analysis in Fig. 5 suggests that the majority of the heat transport is through low-frequency modes with large group velocities and long lifetimes. The calculated JDoS functions for the mixed phases suggest an overall increase in the density of both collision and decay pathways for modes up to ~2 THz compared to the endpoints. For SnS this can be ascribed to chalcogen substitution introducing low-lying optic modes to participate in collision processes, whereas for SnSe the dominant effect is most likely the broadening of the mid-frequency Se phonon bands creating additional scattering channels. In both cases, the small increase in the number of scattering channels would be expected to result in some suppression of heat transport in the alloy, consistent with experimental findings. However, based on the analysis in Fig. 7 we would expect differences in the interaction strengths $P_\lambda$ with composition to have a prominent effect on the heat transport and, as it is not feasible to investigate this further, we therefore treat this finding with caution.

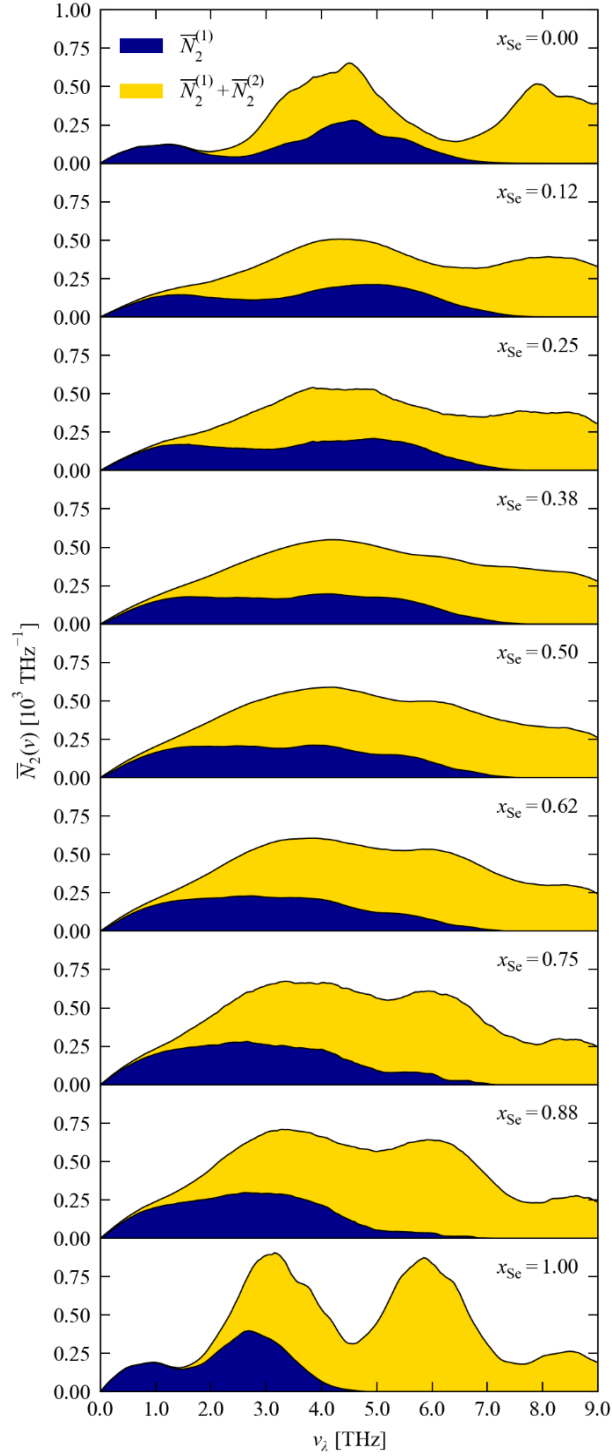

**Figure 8** Averaged two-phonon joint density of states (JDoS) $\overline{N}_2(\nu)$ for Sn(S$_{1-x}$Se$_x$) solid solutions with composition $x_{Se}$ ranging from 0 (SnS) to 1 (SnSe). Each plot shows the JDoS for collision (Type 1 - $\overline{N}_2^{(1)}$) and decay (Type 2 - $\overline{N}_2^{(2)}$) events separately. The data in Fig. 7 for the SnS and SnSe endpoints is included for comparison and was generated using the previous calculations in Refs. [6] and [50]. For the intermediate compositions, the averaging over structures in the solid-solution model was performed based on a formation temperature $T_F$ of 900 K.

**Conclusions**

This work has presented a detailed investigation of the lattice dynamics in *Pnma* Sn(S$_{1-x}$Se$_x$) solid solutions and the impact of the dynamics on the energetics and thermal transport.

Our calculations show that the phonon free energy makes a significant contribution to the thermodynamic partition function, in this case leading to a reduction in the mixing energy and a marked difference in the distribution of occurrence probabilities among individual configurations in the alloy model. Here the qualitative stability of the mixed phases remains unchanged, and the similar electrical properties of the individual structures in the mixed compositions results in negligible changes to the averaged bandgaps. However, given the size of the effects predicted by the calculations, it is very likely that the lattice dynamics could have a significant qualitative impact in other systems, particularly those with energetically-similar competing phases, as has been observed in other work[43,47]. In the case of the Sn$_n$(S$_{1-x}$Se$_x$)$_m$, it is possible that differences in lattice dynamics may bring the metastable rocksalt phase closer to the Sn(S$_{1-x}$Se$_x$) convex hull at intermediate compositions and/or stabilise the sesquisulphide Sn$_2$(S$_{1-x}$Se$_x$)$_3$ relative to the competing mono- and di-chalcogenide phases, but we defer exploration of this to future work.

Comparing the unfolded phonon dispersions and density of states over the range of compositions shows a clear pattern whereby the well-defined bands of high- and mid-frequency S/Se modes in the SnS and SnSe endpoints, respectively, are spread over a wider range of frequencies while the band structure of the low-frequency Sn modes is largely retained. This results in some lowering of the group velocities and produces additional energy-conserving scattering channels in the joint density of states, both of which could in principle reduce the thermal conductivity in the mixed phases. However, comparison of the endpoints suggests that differences in the three-phonon interaction strengths make a significant contribution to the difference in heat transport between them, and it is not practical at present to investigate how these are affected by alloying.

This leads us to close by noting several challenges to this type of study that should be addressed in future work. The practical restriction of calculating the harmonic force constants in the 32-atom alloy supercells is likely restrict the accuracy of calculated phonon spectra and derived properties such as the free energies and group velocities. In general, this issue may be further exacerbated by potentially needing to reduce the plane-wave cutoff and *k*-point sampling in alloy calculations to minimise the computational cost. Secondly, more accurate free energies could be obtained by taking into account the volume expansion at finite temperature using the quasi-harmonic approximation, but to perform QHA calculations for the full set of structures in a complex alloy such as *Pnma* SnS is not currently practical using first-principles methods. Finally, as noted above, our analysis of the thermal conductivity of the SnS and SnSe endpoints indicates that differences in the phonon interaction strengths are likely to be a significant contributor to the differences between the compounds, but to quantify this requires computing the third-order force constants which is again not practical even for a small subset of the structures.

Several routes to address these challenges can be envisaged. The cost of the first-principles calculations could be reduced by using alternatives to plane-wave DFT, for example periodic codes using efficient local-orbital or LCAO approaches[55–57]. Alternatively and/or in addition to this, a sampling approach could be employed to obtain the second- and/or third-order force constants using a smaller set of explicit calculations. The relatively predictable set of chemical environments in an alloy should also make them amenable to modelling with force fields, which could be parameterised against a smaller subset of first-principles calculations in the spirit of the aforementioned sampling approach.

Nonetheless, this study has demonstrated that studying the lattice dynamics of relatively complex solid solutions is within the reach of contemporary first-principles modelling, and has highlighted some of the ways in which changes in the dynamics with composition can influence both the stability of mixed phases and important physical properties for applications such as thermoelectric power.

**Acknowledgements**

JMS gratefully acknowledges the support of a Presidential Fellowship from the University of Manchester and assistance from A. Togo (Kyoto) with using the band-unfolding routines in Phonopy. The majority of the DFT calculations in this work were performed using the UK Archer HPC facility, *via* membership of the UK Materials Chemistry Consortium, which is funded by the UK Engineering and Physical Sciences Research Council (EPSRC; grant nos. EP/L000202 and EP/R029431). The post processing with Phonopy was carried out using the University of Manchester Computational Shared Facility (CSF) HPC cluster, which is supported and maintained by UoM Research IT.

# Lattice dynamics of *Pnma* Sn(S$_{1-x}$Se$_x$) solid solutions: energetics, phonon spectra and thermal transport

**Electronic supporting information**

Jonathan M. Skelton[1]*

[1] *Department of Chemistry, University of Manchester, Oxford Road, Manchester M13 9PL, UK*

* Corresponding author - e-mail jonathan.skelton@manchester.ac.uk

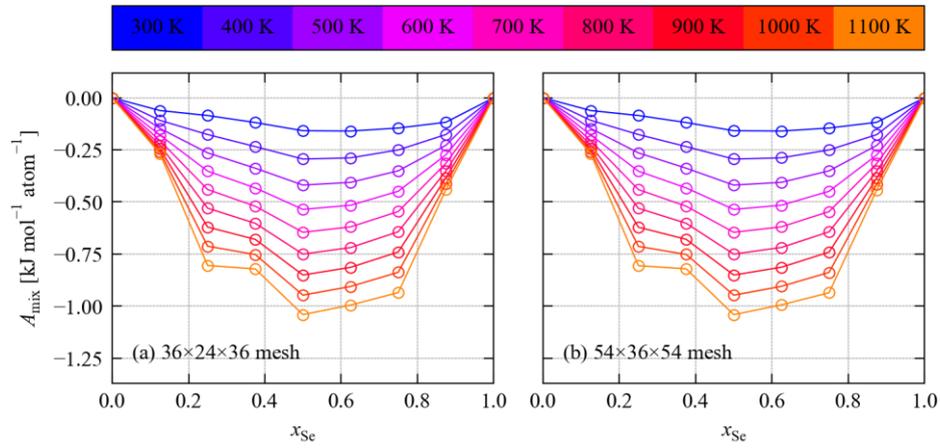

**Figure S1** Calculated mixing free energies $A_\text{mix}$ of *Pnma* Sn(S$_{1-x}$Se$_x$) solid solutions as a function of Se fraction $x_\text{Se}$, obtained using the Helmholtz free energies $A_n$ with the vibrational contributions evaluated using Brillouin zone sampling meshes with 36×24×26 (a) and 54×36×54 subdivisions (b). The data in (a) is the same as that in Fig. 1b in the text. Both plots shows $A_\text{mix}(x_\text{Se})$ for a range of alloy formation temperatures $T_\text{F}$ from 300 - 1100 K, which are denoted by line colours from blue (low $T_\text{F}$) to orange (high $T_\text{F}$).

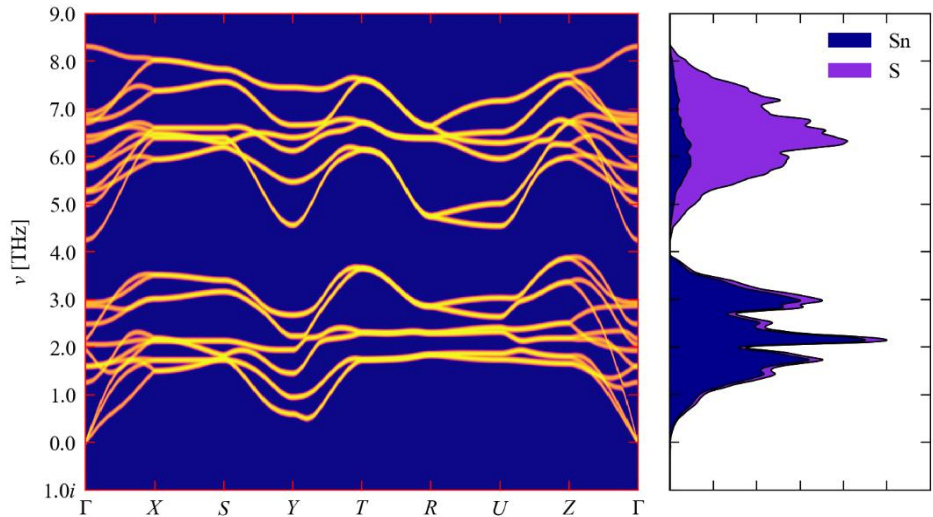

**Figure S2** Simulated phonon spectra of *Pnma* SnS. The left-hand panel shows the phonon dispersion and the right-hand panel shows the density of states $g(\nu)$ (DoS) as a stacked area plot with the projections onto Sn (blue) and S atoms (purple).

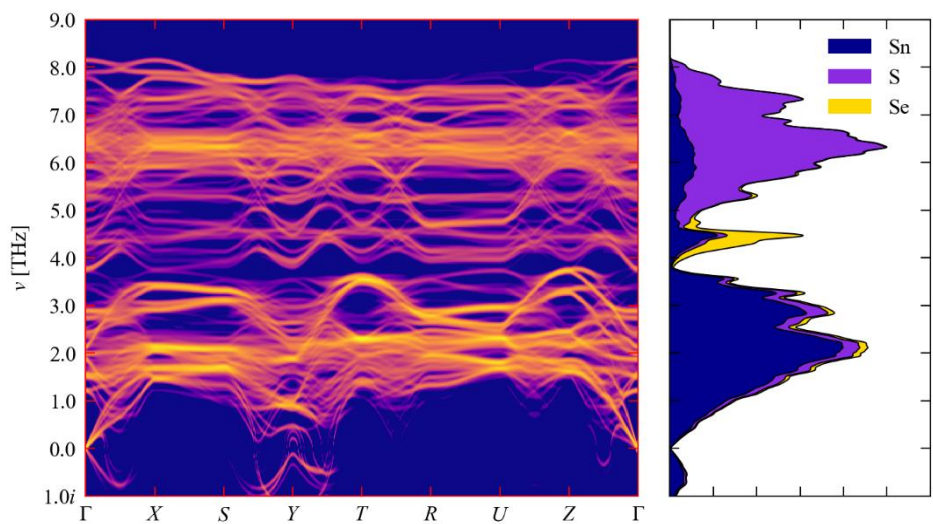

**Figure S3** Simulated phonon spectra of a *Pnma* SnS$_{0.875}$Se$_{0.125}$ solid solution. The left-hand panel shows the phonon dispersion unfolded back to a *Pnma* parent structure. The right-hand panel shows the density of states $g(\nu)$ (DoS) as a stacked area plot with the projections onto Sn (blue), S (purple) and Se atoms (yellow). The averaging over structures in the solid-solution model was performed based on a formation temperature $T_\mathrm{F}$ of 900 K.

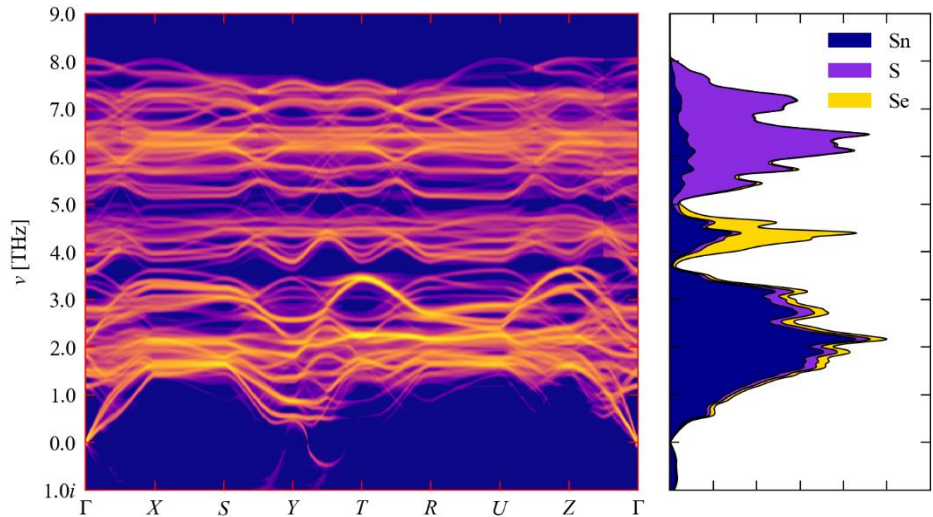

**Figure S4** Simulated phonon spectra of a *Pnma* SnS$_{0.75}$Se$_{0.25}$ solid solution. The left-hand panel shows the phonon dispersion unfolded back to a *Pnma* parent structure. The right-hand panel shows the density of states $g(\nu)$ (DoS) as a stacked area plot with the projections onto Sn (blue), S (purple) and Se atoms (yellow). The averaging over structures in the solid-solution model was performed based on a formation temperature $T_\text{F}$ of 900 K.

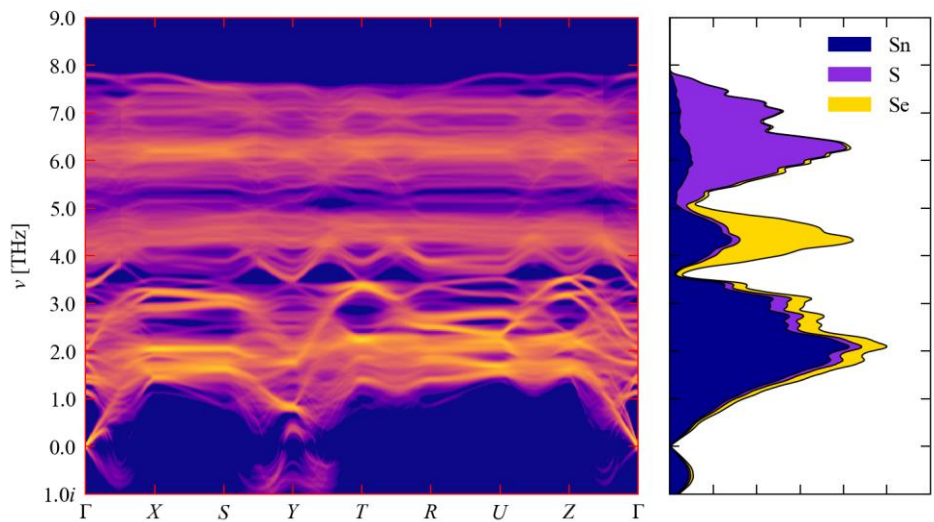

**Figure S5** Simulated phonon spectra of a *Pnma* SnS$_{0.625}$Se$_{0.375}$ solid solution. The left-hand panel shows the phonon dispersion unfolded back to a *Pnma* parent structure. The right-hand panel shows the density of states $g(\nu)$ (DoS) as a stacked area plot with the projections onto Sn (blue), S (purple) and Se atoms (yellow). The averaging over structures in the solid-solution model was performed based on a formation temperature $T_\text{F}$ of 900 K.

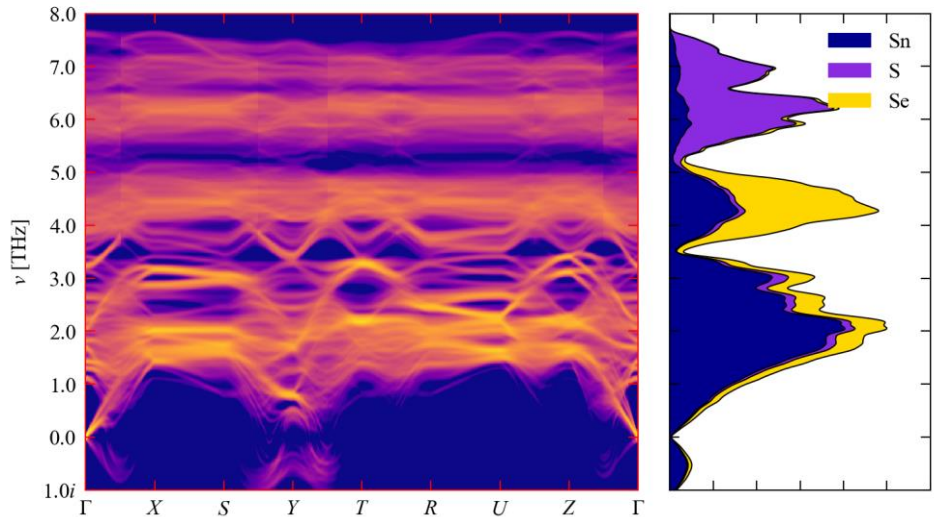

**Figure S6** Simulated phonon spectra of a *Pnma* SnS$_{0.5}$Se$_{0.5}$ solid solution. The left-hand panel shows the phonon dispersion unfolded back to a *Pnma* parent structure. The right-hand panel shows the density of states $g(\nu)$ (DoS) as a stacked area plot with the projections onto Sn (blue), S (purple) and Se atoms (yellow). The averaging over structures in the solid-solution model was performed based on a formation temperature $T_\text{F}$ of 900 K.

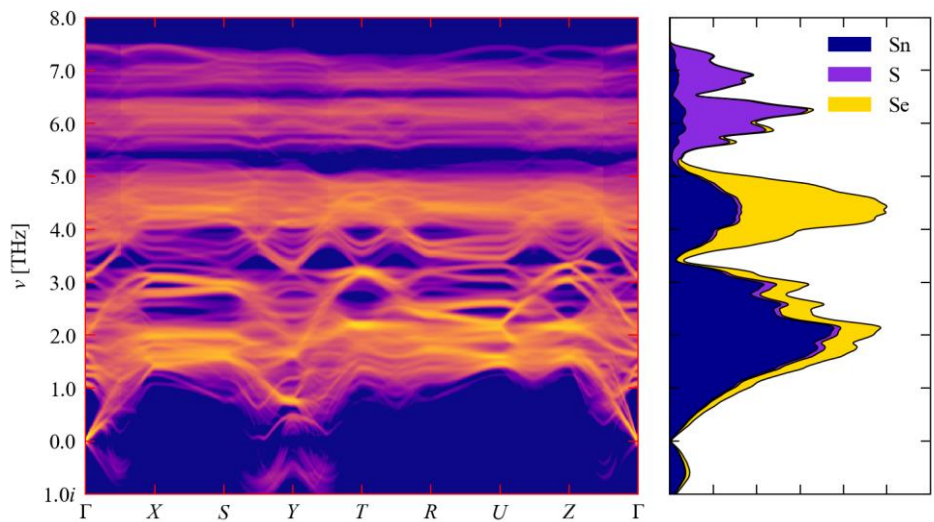

**Figure S7** Simulated phonon spectra of a *Pnma* SnS$_{0.375}$Se$_{0.625}$ solid solution. The left-hand panel shows the phonon dispersion unfolded back to a *Pnma* parent structure. The right-hand panel shows the density of states $g(\nu)$ (DoS) as a stacked area plot with the projections onto Sn (blue), S (purple) and Se atoms (yellow). The averaging over structures in the solid-solution model was performed based on a formation temperature $T_\text{F}$ of 900 K.

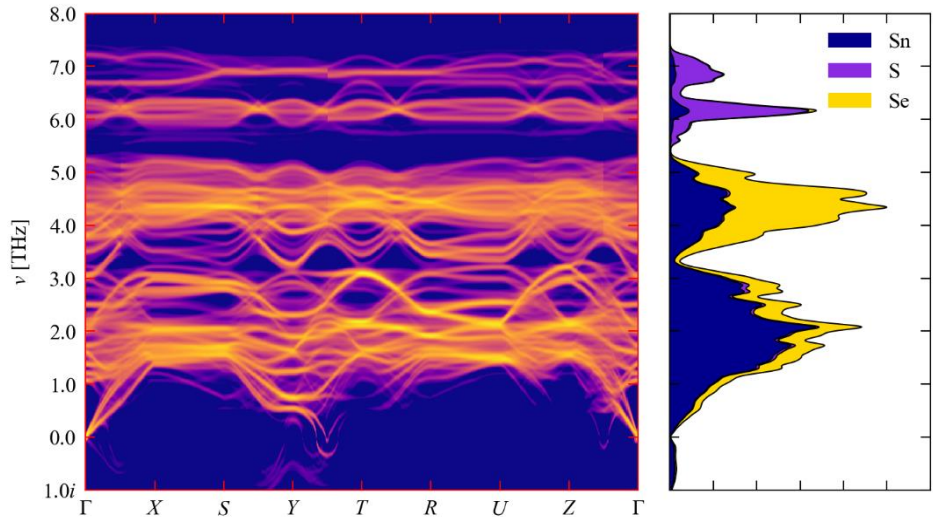

**Figure S8** Simulated phonon spectra of a *Pnma* SnS$_{0.25}$Se$_{0.75}$ solid solution. The left-hand panel shows the phonon dispersion unfolded back to a *Pnma* parent structure. The right-hand panel shows the density of states $g(\nu)$ (DoS) as a stacked area plot with the projections onto Sn (blue), S (purple) and Se atoms (yellow). The averaging over structures in the solid-solution model was performed based on a formation temperature $T_\mathrm{F}$ of 900 K.

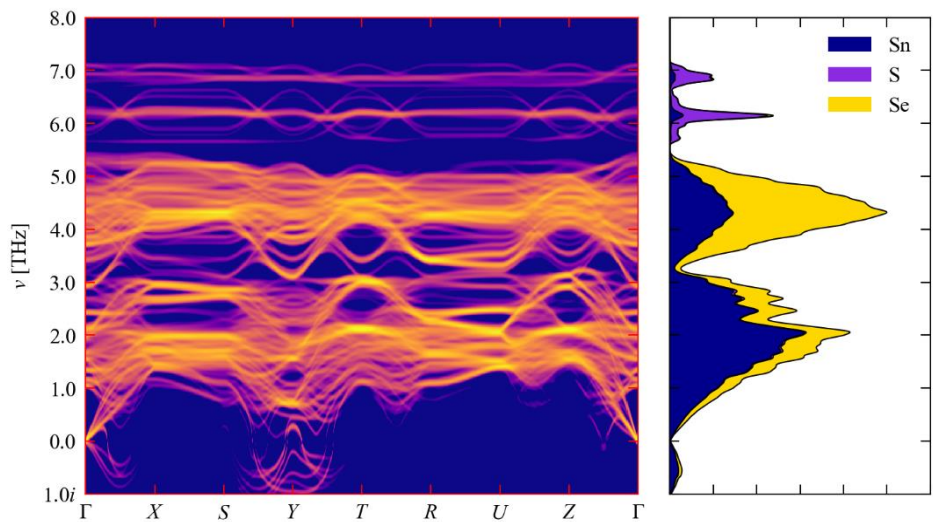

**Figure S9** Simulated phonon spectra of a *Pnma* SnS$_{0.125}$Se$_{0.875}$ solid solution. The left-hand panel shows the phonon dispersion unfolded back to a *Pnma* parent structure. The right-hand panel shows the density of states $g(\nu)$ (DoS) as a stacked area plot with the projections onto Sn (blue), S (purple) and Se atoms (yellow). The averaging over structures in the solid-solution model was performed based on a formation temperature $T_\mathrm{F}$ of 900 K.

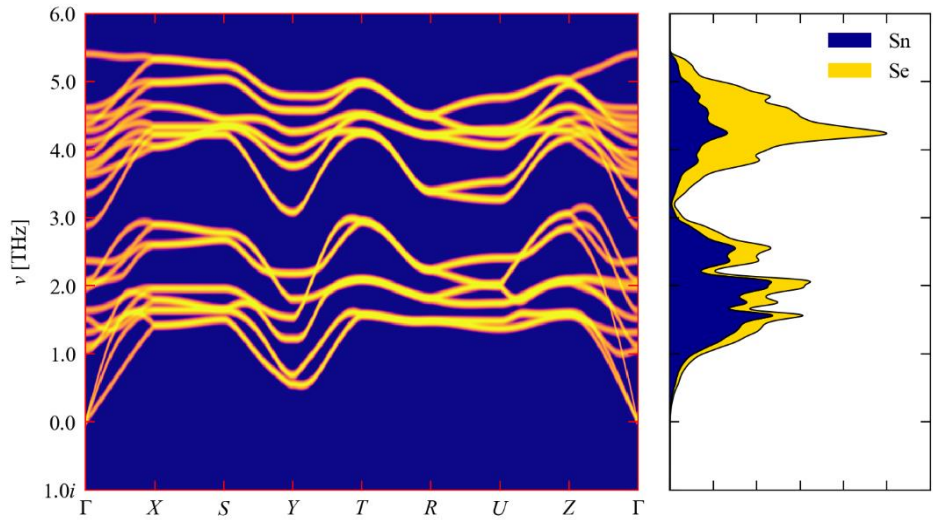

**Figure S10** Simulated phonon spectra of *Pnma* SnSe. The left-hand panel shows the phonon dispersion and the right-hand panel shows the density of states $g(\nu)$ (DoS) as a stacked area plot with the projections onto Sn (blue) and Se atoms (yellow).

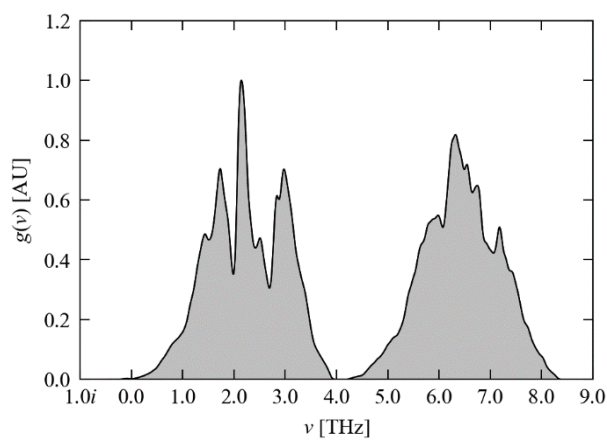

**Figure S11** Simulated phonon density of states of *Pnma* SnS. The DoS was evaluated by interpolating the phonon frequencies onto a regular Γ-centred **q**-point grid with 24×16×24 subdivisions and using Gaussian smearing with a width $\sigma$ = 0.032 THz (FWHM ~ 2.5 cm$^{-1}$) to integrate the Brillouin zone.

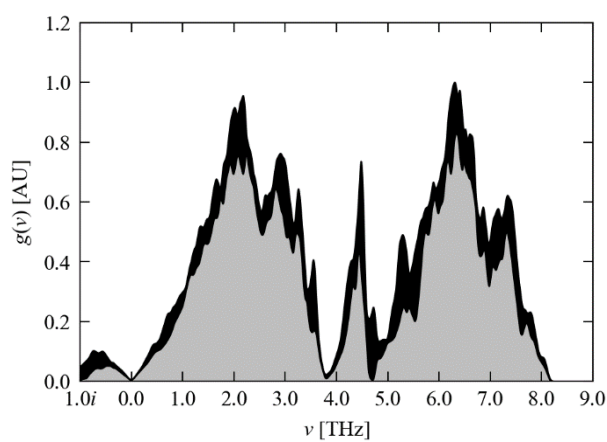

**Figure S12** Simulated phonon density of states of a *Pnma* SnS$_{0.875}$Se$_{0.125}$ solid solution. The grey curve shows the averaged DoS and the shaded black region indicates ± one weighted standard deviation. The DoS was evaluated by interpolating the phonon frequencies onto a regular Γ-centred **q**-point grid with 24×16×24 subdivisions and using Gaussian smearing with a width $\sigma$ = 0.032 THz (FWHM ~ 2.5 cm$^{-1}$) to integrate the Brillouin zone. The averaging over structures in the solid-solution model was performed based on a formation temperature $T_F$ of 900 K.

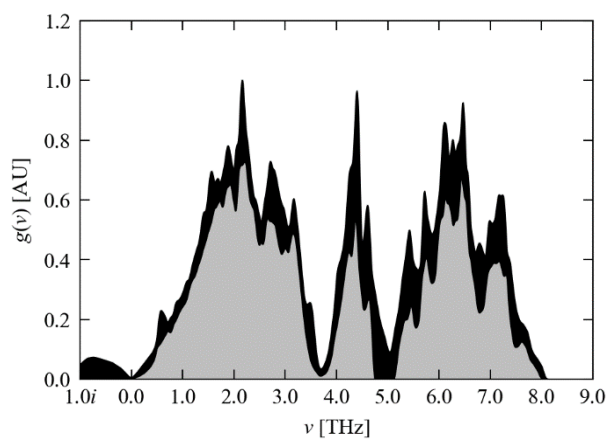

**Figure S13** Simulated phonon density of states of a *Pnma* SnS$_{0.75}$Se$_{0.25}$ solid solution. The grey curve shows the averaged DoS and the shaded black region indicates $\pm$ one weighted standard deviation. The DoS was evaluated by interpolating the phonon frequencies onto a regular Γ-centred **q**-point grid with 24×16×24 subdivisions and using Gaussian smearing with a width $\sigma$ = 0.032 THz (FWHM $\sim$ 2.5 cm$^{-1}$) to integrate the Brillouin zone. The averaging over structures in the solid-solution model was performed based on a formation temperature $T_\mathrm{F}$ of 900 K.

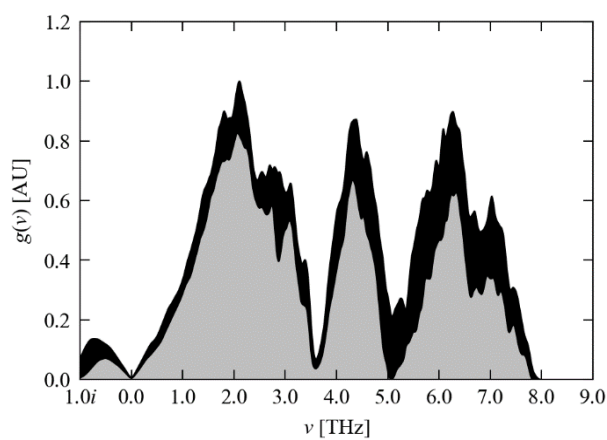

**Figure S14** Simulated phonon density of states of a *Pnma* SnS$_{0.625}$Se$_{0.375}$ solid solution. The grey curve shows the averaged DoS and the shaded black region indicates $\pm$ one weighted standard deviation. The DoS was calcula evaluated ted by interpolating the phonon frequencies onto a regular Γ-centred **q**-point grid with 24×16×24 subdivisions and using Gaussian smearing with a width $\sigma$ = 0.032 THz (FWHM $\sim$ 2.5 cm$^{-1}$) to integrate the Brillouin zone. The averaging over structures in the solid-solution model was performed based on a formation temperature $T_\mathrm{F}$ of 900 K.

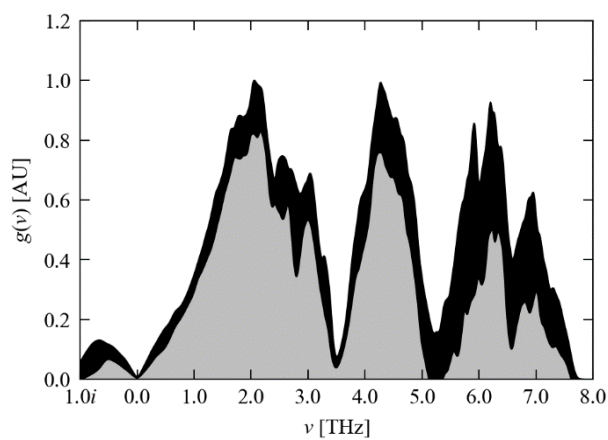

**Figure S15** Simulated phonon density of states of a *Pnma* SnS$_{0.5}$Se$_{0.5}$ solid solution. The grey curve shows the averaged DoS and the shaded black region indicates ± one weighted standard deviation. The DoS was evaluated by interpolating the phonon frequencies onto a regular Γ-centred **q**-point grid with 24×16×24 subdivisions and using Gaussian smearing with a width $\sigma$ = 0.032 THz (FWHM ~ 2.5 cm$^{-1}$) to integrate the Brillouin zone. The averaging over structures in the solid-solution model was performed based on a formation temperature $T_\mathrm{F}$ of 900 K.

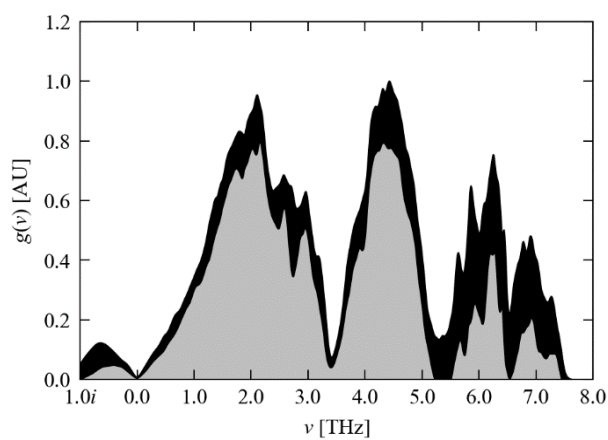

**Figure S16** Simulated phonon density of states of a *Pnma* SnS$_{0.375}$Se$_{0.625}$ solid solution. The grey curve shows the averaged DoS and the shaded black region indicates ± one weighted standard deviation. The DoS was evaluated by interpolating the phonon frequencies onto a regular Γ-centred **q**-point grid with 24×16×24 subdivisions and using Gaussian smearing with a width $\sigma$ = 0.032 THz (FWHM ~ 2.5 cm$^{-1}$) to integrate the Brillouin zone. The averaging over structures in the solid-solution model was performed based on a formation temperature $T_\mathrm{F}$ of 900 K.

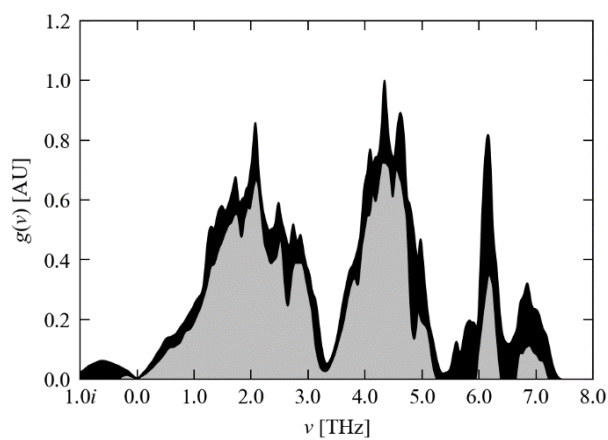

**Figure S17** Simulated phonon density of states of a *Pnma* SnS$_{0.25}$Se$_{0.75}$ solid solution. The grey curve shows the averaged DoS and the shaded black region indicates ± one weighted standard deviation. The DoS was evaluated by interpolating the phonon frequencies onto a regular Γ-centred **q**-point grid with 24×16×24 subdivisions and using Gaussian smearing with a width $\sigma$ = 0.032 THz (FWHM ~ 2.5 cm$^{-1}$) to integrate the Brillouin zone. The averaging over structures in the solid-solution model was performed based on a formation temperature $T_F$ of 900 K.

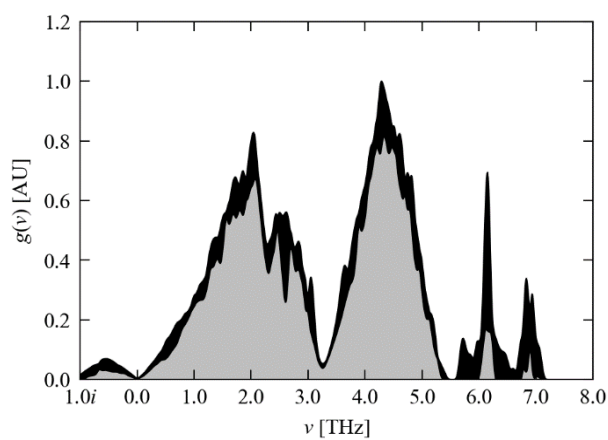

**Figure S18** Simulated phonon density of states of a *Pnma* SnS$_{0.125}$Se$_{0.875}$ solid solution. The grey curve shows the averaged DoS and the shaded black region indicates ± one weighted standard deviation. The DoS was evaluated by interpolating the phonon frequencies onto a regular Γ-centred **q**-point grid with 24×16×24 subdivisions and using Gaussian smearing with a width $\sigma$ = 0.032 THz (FWHM ~ 2.5 cm$^{-1}$) to integrate the Brillouin zone. The averaging over structures in the solid-solution model was performed based on a formation temperature $T_F$ of 900 K.

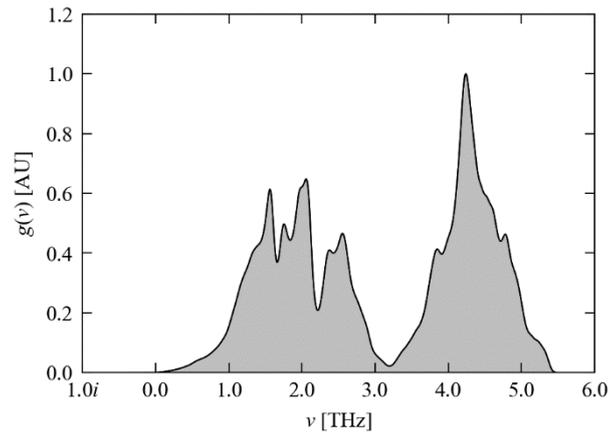

**Figure S19** Simulated phonon density of states of *Pnma* SnSe. The DoS was evaluated by interpolating the phonon frequencies onto a regular Γ-centred **q**-point grid with 24×16×24 subdivisions and using Gaussian smearing with a width $\sigma$ = 0.032 THz (FWHM ∼ 2.5 cm$^{-1}$) to integrate the Brillouin zone.

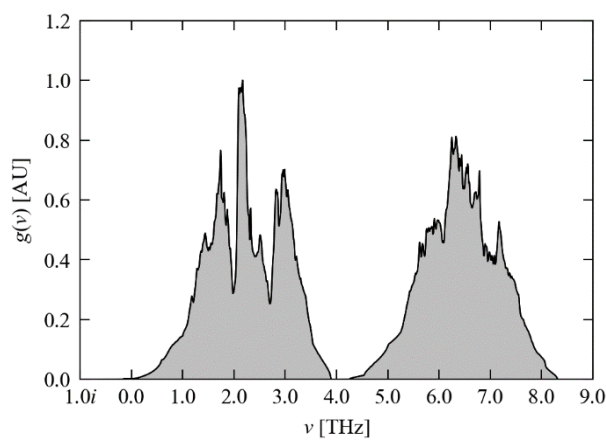

**Figure S20** Simulated phonon density of states of *Pnma* SnS. The DoS was evaluated by interpolating the phonon frequencies onto a regular Γ-centred **q**-point grid with 36×24×36 subdivisions and using the linear tetrahedron method to integrate the Brillouin zone.

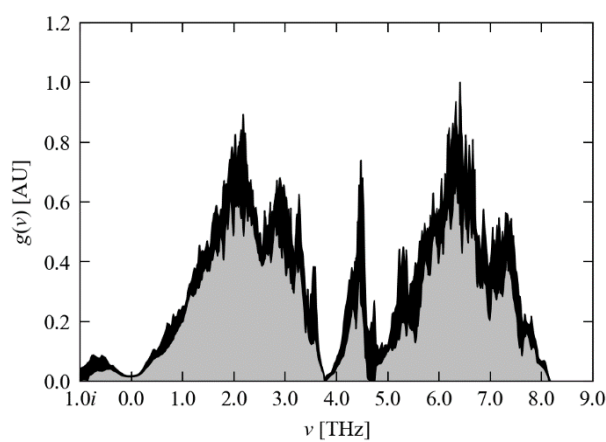

**Figure S21** Simulated phonon density of states of a *Pnma* SnS$_{0.875}$Se$_{0.125}$ solid solution. The grey curve shows the averaged DoS and the shaded black region indicates ± one weighted standard deviation. The DoS was evaluated by interpolating the phonon frequencies onto a regular Γ-centred **q**-point grid with 36×24×36 subdivisions and using the linear tetrahedron method to integrate the Brillouin zone. The averaging over structures in the solid-solution model was performed based on a formation temperature $T_\mathrm{F}$ of 900 K.

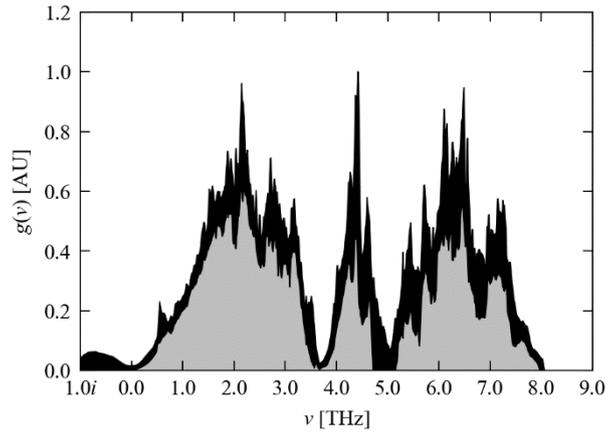

**Figure S22** Simulated phonon density of states of a *Pnma* SnS$_{0.75}$Se$_{0.25}$ solid solution. The grey curve shows the averaged DoS and the shaded black region indicates ± one weighted standard deviation. The DoS was evaluated by interpolating the phonon frequencies onto a regular Γ-centred **q**-point grid with 36×24×36 subdivisions and using the linear tetrahedron method to integrate the Brillouin zone. The averaging over structures in the solid-solution model was performed based on a formation temperature $T_\mathrm{F}$ of 900 K.

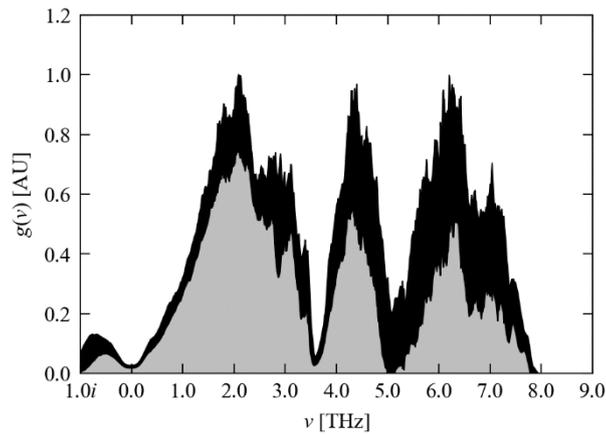

**Figure S23** Simulated phonon density of states of a *Pnma* SnS$_{0.625}$Se$_{0.375}$ solid solution. The grey curve shows the averaged DoS and the shaded black region indicates ± one weighted standard deviation. The DoS was evaluated by interpolating the phonon frequencies onto a regular Γ-centred **q**-point grid with 36×24×36 subdivisions and using the linear tetrahedron method to integrate the Brillouin zone. The averaging over structures in the solid-solution model was performed based on a formation temperature $T_\mathrm{F}$ of 900 K.

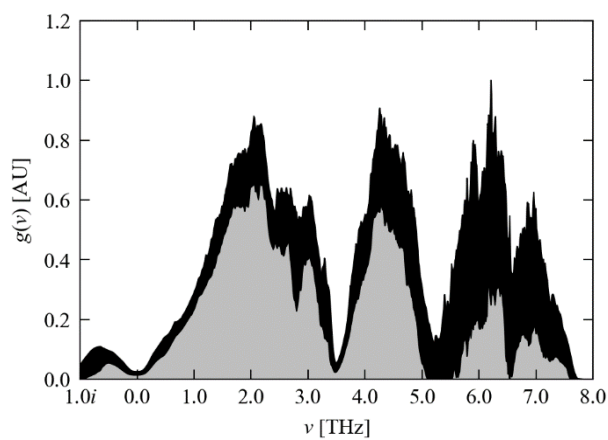

**Figure S24** Simulated phonon density of states of a *Pnma* SnS$_{0.5}$Se$_{0.5}$ solid solution. The grey curve shows the averaged DoS and the shaded black region indicates ± one weighted standard deviation. The DoS was evaluated by interpolating the phonon frequencies onto a regular Γ-centred **q**-point grid with 36×24×36 subdivisions and using the linear tetrahedron method to integrate the Brillouin zone. The averaging over structures in the solid-solution model was performed based on a formation temperature $T_\mathrm{F}$ of 900 K.

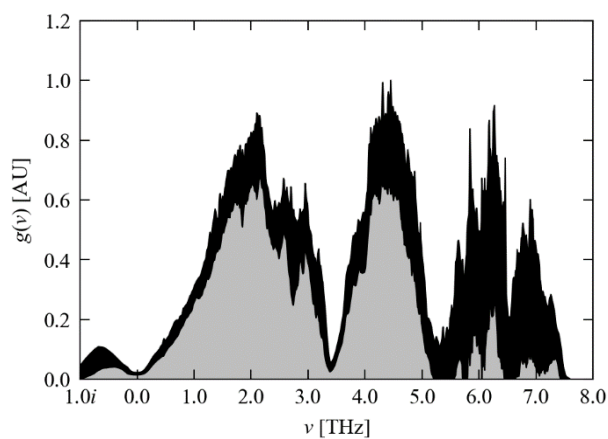

**Figure S25** Simulated phonon density of states of a *Pnma* SnS$_{0.375}$Se$_{0.625}$ solid solution. The grey curve shows the averaged DoS and the shaded black region indicates ± one weighted standard deviation. The DoS was evaluated by interpolating the phonon frequencies onto a regular Γ-centred **q**-point grid with 36×24×36 subdivisions and using the linear tetrahedron method to integrate the Brillouin zone. The averaging over structures in the solid-solution model was performed based on a formation temperature $T_\mathrm{F}$ of 900 K.

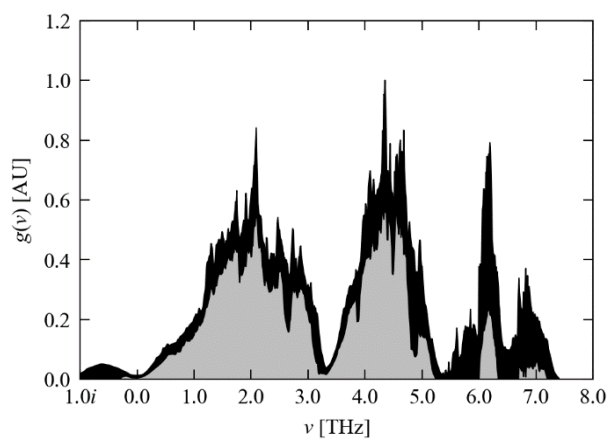

**Figure S26** Simulated phonon density of states of a *Pnma* SnS$_{0.25}$Se$_{0.75}$ solid solution. The grey curve shows the averaged DoS and the shaded black region indicates ± one weighted standard deviation. The DoS was evaluated by interpolating the phonon frequencies onto a regular Γ-centred **q**-point grid with 36×24×36 subdivisions and using the linear tetrahedron method to integrate the Brillouin zone. The averaging over structures in the solid-solution model was performed based on a formation temperature $T_\mathrm{F}$ of 900 K.

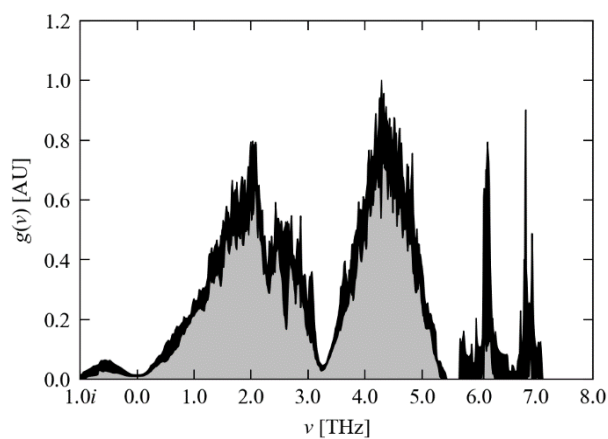

**Figure S27** Simulated phonon density of states of a *Pnma* SnS$_{0.125}$Se$_{0.875}$ solid solution. The grey curve shows the averaged DoS and the shaded black region indicates ± one weighted standard deviation. The DoS was evaluated by interpolating the phonon frequencies onto a regular Γ-centred **q**-point grid with 36×24×36 subdivisions and using the linear tetrahedron method to integrate the Brillouin zone. The averaging over structures in the solid-solution model was performed based on a formation temperature $T_\mathrm{F}$ of 900 K.

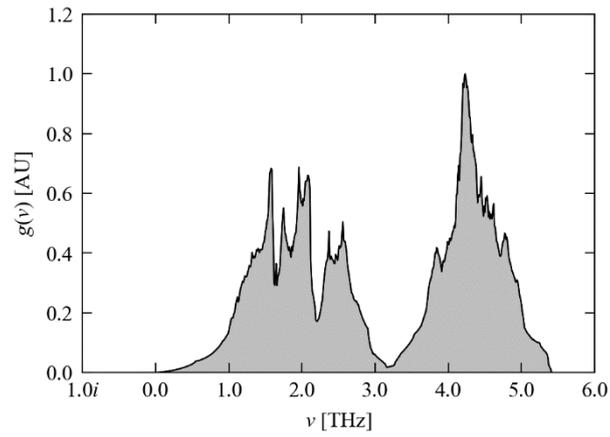

**Figure S28** Simulated phonon density of states of *Pnma* SnSe. The DoS was evaluated by interpolating the phonon frequencies onto a regular Γ-centred **q**-point grid with 36×24×36 subdivisions and using the linear tetrahedron method to integrate the Brillouin zone.

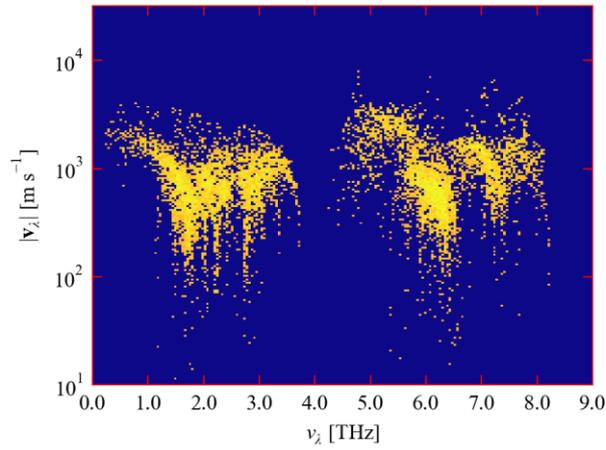

**Figure S29** Simulated frequency spectrum of the group velocity norms $|v_\lambda|$ in *Pnma* SnS. The colour scale indicates the density of modes and runs from red (low density) to yellow (high density). As in Fig. 5 in the text, this data was taken from the previous calculation in Ref. [1].

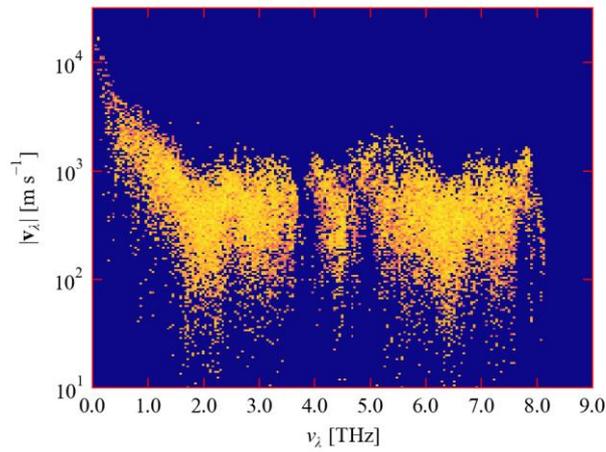

**Figure S30** Simulated frequency spectrum of the group velocity norms $|v_\lambda|$ in a *Pnma* SnS$_{0.875}$Se$_{0.125}$ solid solution. The colour scale indicates the density of modes and runs from red (low density) to yellow (high density). The averaging over structures in the solid-solution model was performed based on a formation temperature $T_F$ of 900 K.

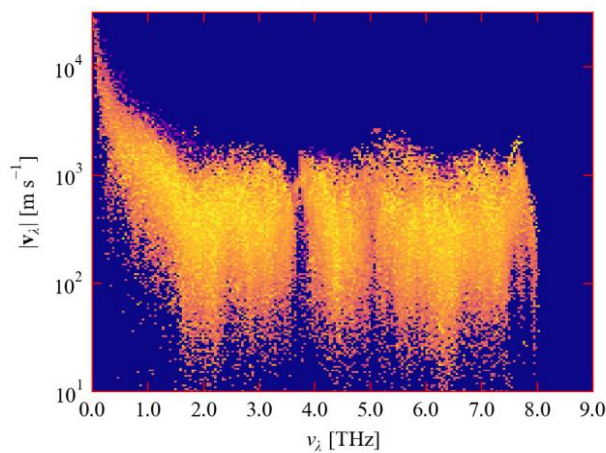

**Figure S31** Simulated frequency spectrum of the group velocity norms $|v_\lambda|$ in a *Pnma* SnS$_{0.75}$Se$_{0.25}$ solid solution. The colour scale indicates the density of modes and runs from red (low density) to yellow (high density). The averaging over structures in the solid-solution model was performed based on a formation temperature $T_\mathrm{F}$ of 900 K.

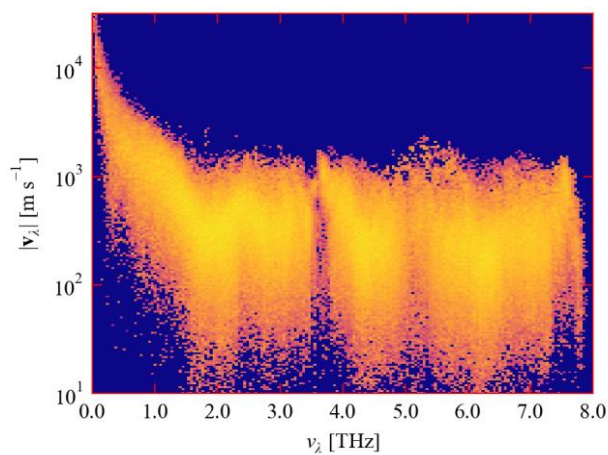

**Figure S32** Simulated frequency spectrum of the group velocity norms $|v_\lambda|$ in a *Pnma* SnS$_{0.625}$Se$_{0.375}$ solid solution. The colour scale indicates the density of modes and runs from red (low density) to yellow (high density). The averaging over structures in the solid-solution model was performed based on a formation temperature $T_\mathrm{F}$ of 900 K.

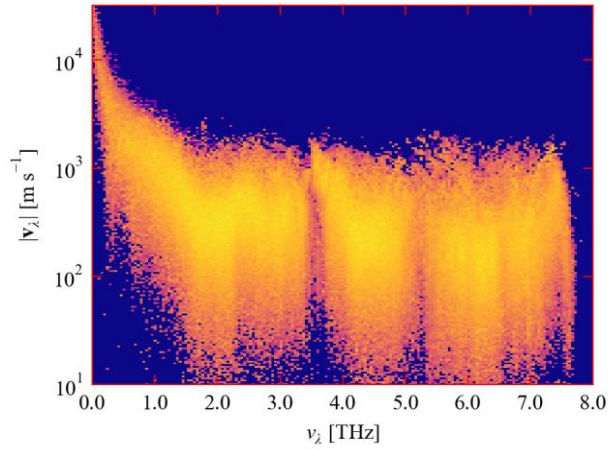

**Figure S33** Simulated frequency spectrum of the group velocity norms $|v_\lambda|$ in a *Pnma* SnS$_{0.5}$Se$_{0.5}$ solid solution. The colour scale indicates the density of modes and runs from red (low density) to yellow (high density). The averaging over structures in the solid-solution model was performed based on a formation temperature $T_\mathrm{F}$ of 900 K.

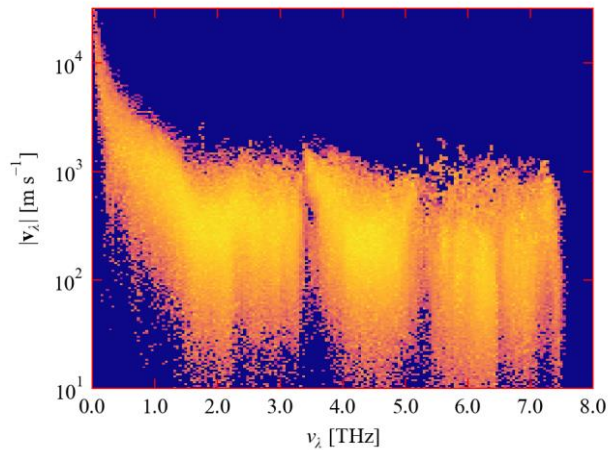

**Figure S34** Simulated frequency spectrum of the group velocity norms $|v_\lambda|$ in a *Pnma* SnS$_{0.375}$Se$_{0.625}$ solid solution. The colour scale indicates the density of modes and runs from red (low density) to yellow (high density). The averaging over structures in the solid-solution model was performed based on a formation temperature $T_\mathrm{F}$ of 900 K.

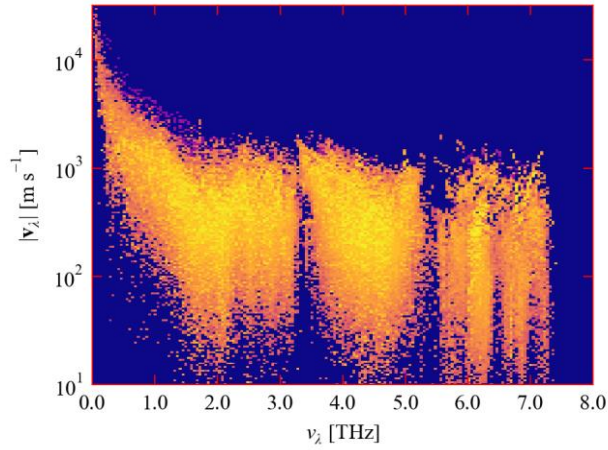

**Figure S35** Simulated frequency spectrum of the group velocity norms $|v_\lambda|$ in a *Pnma* $SnS_{0.25}Se_{0.75}$ solid solution. The colour scale indicates the density of modes and runs from red (low density) to yellow (high density). The averaging over structures in the solid-solution model was performed based on a formation temperature $T_F$ of 900 K.

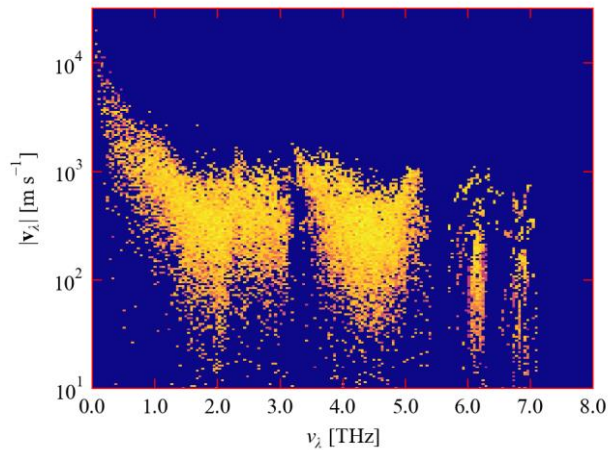

**Figure S36** Simulated frequency spectrum of the group velocities in a *Pnma* $SnS_{0.125}Se_{0.875}$ solid solution. The colour scale indicates the density of modes and runs from red (low density) to yellow (high density). The averaging over structures in the solid-solution model was performed based on a formation temperature $T_F$ of 900 K.

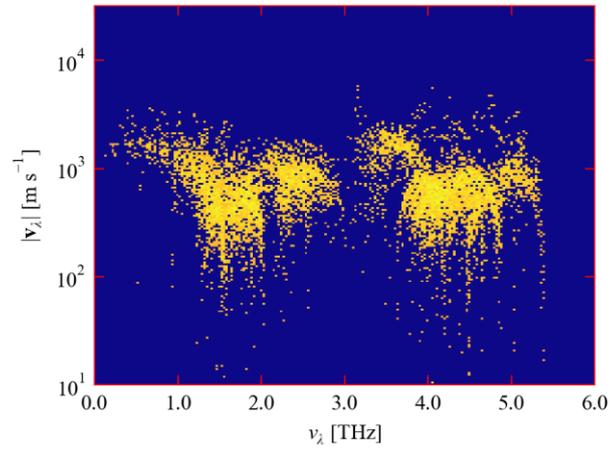

**Figure S37** Simulated frequency spectrum of the group velocity norms $|v_\lambda|$ in *Pnma* SnSe. The colour scale indicates the density of modes and runs from red (low density) to yellow (high density). As in Fig. 5 in the text, this data was taken from the previous calculation in Ref. [2].